\documentclass[a4paper,fleqn,usenatbib]{mnras}

\usepackage{newtxtext,newtxmath}

\usepackage[T1]{fontenc}
\usepackage{ae,aecompl}

\usepackage{graphicx}
\usepackage{amsmath}	
\usepackage{amssymb}	

\defcitealias{2014MNRAS.443.3463S}{SRNS14}

\begin{document}
\title[Supernovae to Superbubbles]{How multiple supernovae overlap to form superbubbles}
\author[Yadav, Mukherjee, Sharma, \& Nath]{Naveen Yadav$^{1}$\thanks{{naveen.phys@gmail.com}},
Dipanjan Mukherjee$^{2}$\thanks{{dipanjan.mukherjee@anu.edu.au}},
Prateek Sharma$^{1}$\thanks{{prateek@physics.iisc.ernet.in}}
and Biman B. Nath$^{3}$\\
$^{1}$Department of Physics \& Joint Astronomy Programme, Indian Institute of Science, Bangalore,
India -560012. \\
$^{2}$Research School of Astronomy \& Astrophysics, Mount Stromlo Observatory, ACT 2611, Australia.\\
$^{3}$Raman Research Institute, Sadashiva Nagar, Bangalore, India -560080.}

\maketitle

\begin{abstract}
We explore the formation of superbubbles through energy deposition by multiple supernovae (SNe) in
a uniform medium. We use total energy conserving, 3-D hydrodynamic simulations to study how SNe
correlated in space and time create superbubbles. While
isolated SNe fizzle out completely by $\sim 1$ Myr due to radiative losses, for a realistic cluster size it is likely
that subsequent SNe go off within the hot/dilute bubble and sustain the shock till the cluster lifetime.
For realistic cluster sizes, we find
that the bubble remains overpressured only if, for a given $n_{g0}$, $N_{\rm OB}$ is sufficiently
large. While most of the input energy is still lost radiatively, superbubbles can retain up to
$\sim 5-10\%$ of the input energy in form of kinetic+thermal energy till 10 Myr for ISM density $n_{g0} \approx 1$ cm$^{-3}$.
We find that the mechanical efficiency decreases for higher densities ($\eta_{\rm mech} \propto n_{g0}^{-2/3}$). We compare the radii and velocities
of simulated supershells with observations and the classical adiabatic model. Our simulations show that the superbubbles retain only
$\lesssim 10\%$ of the injected energy, thereby explaining the observed smaller size and slower expansion of supershells.
We also confirm that a sufficiently large ($\gtrsim 10^4$) number of SNe is required to go off in order to
create a steady wind with a stable termination shock within the superbubble. We show that the mechanical efficiency increases 
with increasing resolution, and that explicit diffusion is required to obtain converged results.
\end{abstract}


\section{Introduction}
H\footnotesize{I} holes, shells, rings, expanding cavities, galactic chimneys, and filaments are
ubiquitous structures which are embedded in the large scale gas distribution of a galaxy.
\citet{1979ApJ...229..533H} identified large cavities in the local interstellar medium (ISM) with
energy requirement of $\gtrsim 3\times 10^{52}\ \rm erg$ as supershells. Our solar system is itself
embedded in such a cavity (radius $\sim 100\ \rm pc$) filled with hot ($\sim 10^6\ \rm K$) and
tenuous ($n\sim 5\times 10^{-3}\ \rm cm^{-3}$) plasma \citep{1977ApJ...217L..87S,1983ApJ...269..107M}
known as the local hot bubble (LHB).

When the size of a superbubble becomes comparable to the galactic HI scale height, it may break out
of the galactic disk if the shell is sufficiently fast \citep[e.g.,][]{1988ApJ...324..776M,2013MNRAS.434.3572R}
and inject energy and metals into the galactic halo. The widely accepted model of galaxy-scale superwinds
involves injection of mechanical energy by massive stars in the form of radiation ($L_{\star}$), stellar
winds ($L_{\rm w}$) and supernova (SN) explosions ($E_{\rm SN}\sim 10^{51}\ \rm erg$). Clearly, such large
cavities cannot be created by either the wind from a single massive star or by the supernova explosion
of a single star. Further it is known from observations of O-type stars in the Galaxy that $\sim 70\%$ of
them are associated with clusters and OB associations and a very small fraction of the known O-stars are
isolated \citep{2008ASPC..387..415C}. Out of the remaining $30\%$, more than
one-third are runaway stars which have been ejected in close gravitational encounters \citep{1987ApJS...64..545G}.
Hence the most plausible
mechanism for the formation of large superbubbles is quasi-continuous energy injection from multiple stars.
The expanding shells of each individual star/SN merge to form a large scale bubble
known as a superbubble.

\citet{1968ApL.....2...97P,1972SvA....15..708A} studied the interaction of a strong stellar wind with the
interstellar medium (ISM). The circumstellar shell enters the snowplow phase when the radiative cooling
timescale for the swept gas becomes equal to the dynamical age of the shell. \citet{1977ApJ...218..377W}
calculated the detailed structure for interaction of a strong stellar wind with the interstellar medium.
\citet{1975ApJ...200L.107C} obtained a solution for the case of continuous energy injection (at a point)
inside a homogeneous medium by a stellar wind ($L_{\rm w}=\dot{M}v_{\rm w}^2/2$) in the absence of radiative
energy losses and found the presence of a transition region dominated by thermal conduction between the cold
outer layer of the shell (shocked ISM) and the hot inner layer of the shell (shocked stellar wind).
\citet{1977ApJ...218..377W} analytically calculated detailed structure of the bubble in various phases of
evolution, including the effects of radiative cooling. \citet{1987ApJ...317..190M} highlighted that the
stellar initial mass function and stellar ages are such that the impact of mechanical energy input from
supernovae (SNe) within a star cluster can be well modeled as a constant luminosity driven superbubble.

\citet{1985Natur.317...44C} obtained the steady wind solution driven by a constant rate of mass and
thermal energy injection within a small spherical volume. Their solution is subsonic within the injection
radius, and beyond that reaches a constant supersonic speed. They applied their wind solution to understand
the observations of the galactic outflow in M82. \citet{1981Ap&SS..78..273T} performed 1-D calculations in
a medium with constant particle density. In their calculations all explosions occur at the same point in
space sequentially inside the cavity created by previous SNe. \citet{2015MNRAS.446.1703V} have
recently carried out 3-D simulations in which SNe are uniformly distributed throughout the simulation
box. \citet{2012MNRAS.419..465D} studied the concordance of supernovae feedback methods based on 
thermal energy deposition and kinetic energy deposition.
\citet{2014MNRAS.443.3463S} (hereafter, \citetalias{2014MNRAS.443.3463S}) show that isolated supernova,
in typical ISM conditions, lose almost all
their mechanical energy by radiative losses by $\lesssim 0.1\ \rm Myr$, whereas a sequence of explosions
occurring inside the cavity blown by previous SNe can retain up to $\sim 40\%$ of the injected
mechanical energy for few tens of Myr (of order the galactic dynamical time $\sim 50$ Myr).
\citet{2013A&A...550A..49K,2014A&A...566A..94K} have studied the evolution of interacting
interstellar bubbles of three massive stars in a uniform medium. Their key finding is that 
a larger fraction of energy is retained in the ISM for more closely packed stars. 
The hot bubble mostly emits in soft X-rays below $1.0\ \rm keV$.

Understanding the impact of massive stars, via their radiation, winds, and SNe, on the ISM is
essential for star and galaxy formation. Observed star formation is inefficient, both locally on molecular
cloud scales (e.g., \citealt{2007ApJ...654..304K}) and globally on galactic scales
(e.g., \citealt{2003MNRAS.341...54K}). Supersonic turbulence, magnetic fields, radiative, photoionization
and jet feedback from massive stars, etc. are invoked to explain the inefficiency of star formation on
molecular cloud scales (\citealt{2014prpl.conf..243K} and references therein). Because of several complex
processes involved, there is no consensus on the relative contribution of these different mechanisms acting
on molecular cloud scales. The situation is slightly better on galactic scales ($\gtrsim 1$ kpc)
at which thermal supernova feedback seems to be the dominant mechanism for regulating star formation (e.g.,
\citealt{2004ApJ...606..829S,2005A&A...436..585D,2006ApJ...653.1266J,2013MNRAS.429.1922C,2014A&A...570A..81H,2015ApJ...814....4L}).

It is well recognized that isolated SNe suffer catastrophic cooling losses in high density clouds in which
they are born (e.g., \citealt{1998ApJ...500...95T}). In this case, almost all of the injected energy is lost rather
than coupling to the ISM, especially over global dynamical timescales ($\sim 10$s of Myr). Even when SNe
coalesce before each of them suffer radiative losses (i.e., if supernova rate density is high enough), they only
retain $\sim 10\%$ of the injected energy (\citealt{2012MNRAS.427.1219S}). Even such a small efficiency of mechanical
energy coupling to the ISM appears more than enough to significantly suppress star formation
on global scales for Milky Way and lower mass galaxies (e.g., \citealt{2000MNRAS.317..697E}; red dot-dashed
line in Fig. 4 of \citealt{2012MNRAS.427.1219S}).

Shocks generated by supersonic turbulence (expected within the dense shell) enhance density perturbations and gravitational
instability {\it locally} (e.g., \citealt{1987ApJ...317..190M}), but turbulence and magnetic fields in the dense shell, in all likelihood,
prevent efficient {\it global} star formation (e.g., \citealt{1992ApJ...390L..17S,2004RvMP...76..125M}). Since turbulence
can only be faithfully captured in 3-D, it is necessary to study the ISM using 3-D simulations.

The problem of star-ISM interaction involves complex chemical, ionization/recombination, thermal, and dynamical
processes, and it is necessary to begin with understanding the most important processes in some detail. 
Multi-physics simulations (including gravity, chemistry, photoelectric heating, molecular physics 
and supernovae feedback) of ISM have been done by many authors 
(e.g., \citealt{2015MNRAS.449.1057G,2015MNRAS.450..504M,2015MNRAS.451.2757W,2015MNRAS.454..238W}).
In this paper
we ignore all these processes except for idealized dynamical and thermal processes associated with SNe
resulting from the death of massive stars. We also ignore magnetic fields and thermal conduction, which can greatly
modify the structures with large temperature gradients (e.g., Fig. 9 of \citetalias{2014MNRAS.443.3463S}).
We only consider the hot and warm phases of the ISM by turning off cooling below $10^4$ K, corresponding to the thermally stable
warm neutral medium of the ISM. We do not consider the denser cold neutral phase because: (i) the stable cold phase
exists globally only for a large enough ISM pressure, and hence is unlikely to be present in substantial amount in galaxies
less massive than Milky Way (\citealt{1995ApJ...443..152W}); (ii) our focus is on feedback at scales larger than
molecular clouds, and we assume that a good fraction of supernova energy is able to leak out (aided by low density
channels formed due to stellar winds and radiation) into the more uniformly spread and geometrically thicker warm
neutral disk. Thus, this paper is a generalization of 1-D simulations of \citetalias{2014MNRAS.443.3463S}, with a
realistic spatial distribution of SNe in 3-D. Unlike that work, we also use a total energy conserving code so
that the value of mechanical efficiency is more accurate.

In this paper we study the formation of superbubbles using idealized 3-D hydrodynamic numerical
simulations of SNe exploding in an initially homogeneous, isotropic ISM. Ours are among the
highest resolution uniform-grid 3-D simulations of their kind. In section \ref{sec:setup} we describe the
physical setup and numerical simulations. In section \ref{sec:fiducial} we describe the key results from
our simulations. Section \ref{sec:discussion} discusses analytic estimates and implications of our work.
In section \ref{sec:conclusions} we conclude.
\section{Physical setup}\label{sec:setup}
We choose an idealized physical setup of a uniform ISM at $10^4$ K, corresponding to the warm neutral
medium (WNM) maintained in thermal balance by photoelectric/photoionization and cosmic ray heating
(\citealt{1995ApJ...443..152W}). The Milky Way Giant Molecular Clouds (GMCs) have gas $(\rm H_2)$ densities 
ranging from  $10-1000\ \rm cm^{−3}$ and mean size around $\approx 10 - 20\ \rm pc$, as shown in \citet{2010ApJ...723..492R}. 
Our scales of interest are much bigger 
($\gtrsim 100\ \rm pc$), corresponding to the WNM. Thermal energy is injected 
by SNe going off at random locations
inside a spherical `star cluster' and plasma is allowed to cool due to free-free and line emission till $10^4$ K.
The key aim is to study the dynamics and thermodynamics of SNe coalescing in the WNM, and to
study the conditions for the formation of overpressured superbubbles.
\subsection{Simulation Setup}
We solve the hydrodynamic equations for the evolution of density, velocity and pressure in 3-D Cartesian
coordinates using the static grid version of the finite volume, conservative, Godunov Eulerian code  \texttt{PLUTO}
\citep{2007ApJS..170..228M}.
The mass and energy injected due to SNe are added as source terms. The grid spacing is taken to be
uniform in $x$, $y$ and $z$ directions. We numerically solve the following equations:
\begin{eqnarray}
&&\frac{\partial \rho}{\partial t} + \mathbf{v}\cdot\nabla\rho +
\rho\nabla\cdot \mathbf{v} = \dot{\rho}_{\rm SN}(t,\mathbf{x}) \label{eq.rho}, \\
&&\frac{\partial\mathbf{v}}{\partial t} + \mathbf{v}\cdot\nabla\mathbf{v}
+ \frac{1}{\rho}\nabla p = 0 \label{eq.vel}, \\
&&\frac{\partial p}{\partial t} + \mathbf{v}\cdot\nabla p
+ \rho c^2_s\nabla\cdot \mathbf{v} = (\gamma-1)\{\dot{e}_{\rm SN}(t,\mathbf{x}) -\dot{e}_{\rm rad}(t,\mathbf{x})\},
\label{eq.energy}
\end{eqnarray}
where symbols have their usual meanings, $c_s=(\gamma p/\rho)^{1/2}$ is the sound speed,  $\dot{\rho}_{\rm SN}$
is the mass density source term, $\dot{e}_{\rm SN}$ is the thermal energy source term mimicking supernova feedback
(see section.~\ref{sec:eng-inj}), $\dot{e}_{\rm rad}\equiv n_e n_i \Lambda[T]$ ($n_e$ is electron number density,
$n_i$ is ion number density and $\Lambda[T]$ is the temperature-dependent cooling function) is the rate of energy
loss per unit volume due to radiative cooling. We use the ideal gas equation
\begin{equation}
 \rho\epsilon = \frac{p}{(\gamma-1)}
 \label{eqn:thermal-eng}
\end{equation}
with $\gamma=5/3$ ($\epsilon$ is internal energy per unit mass).

\texttt{PLUTO} solves the system of conservation laws which can be written as \begin{equation}
 \frac{\partial \mathbf{u}}{\partial t } = -\nabla\cdot \mathbf{\Pi}
 + \mathbf{S}, \label{eqn:cons-law}
\end{equation}
where $\mathbf{u}$ is a vector of conserved quantities, $\mathbf{\Pi}$ is the flux tensor and
$\mathbf{S}$ is the source term. The system of equations is integrated using finite volume methods. The
temporal evolution of Eq.~\ref{eqn:cons-law} is carried by explicit methods and the time step is limited by the
Courant-Friedrichs-Lewy (CFL; \citealt{1928MatAn.100...32C}) condition.
The code implements time-dependent optically thin cooling ($\dot{e}_{\rm rad}$ in Eq. \ref{eq.energy})
and the source terms ($\dot{\rho}_{\rm SN}$ in Eq. \ref{eq.rho} and $\dot{e}_{\rm SN}$ in Eq. \ref{eq.energy})
via operator splitting. Our results
are unaffected by boundary conditions because we ensure that our box-size is large enough such that the outer
shock is sufficiently inside the computational domain. We use the HLLC Riemann solver \citep{1994ShWav...4...25T}.
The solution is advanced in time using a second order Runge-Kutta (RK2) scheme and a total variation
diminishing (TVD) linear interpolation is used. The CFL number is set to $0.3$ for numerical stability.
The computational domain is initialized with an interstellar medium (ISM) of uniform density ($n_{g0}$),
with a mean molecular weight per particle $\mu = 0.603$ (mean molecular weight per electron $\mu_e=1.156$)
and solar metallicity at a temperature of $10^{4} \rm \ K$.

We have used the cooling module of \texttt{PLUTO} with the solar metallicity cooling
table of \citealt{1993ApJS...88..253S}).The cooling function is set to zero below $10^{4} \ \rm K$.
We do not include self-gravity, disk stratification, magnetic fields, and any form of gas heating (except by thermal
energy injection due to SNe) in our simulations.

We have two types of simulation setups:
\begin{itemize}
\item \textbf{Full box:} The full box simulations have a computational domain extending from $-L$ to $+L$ in all
three directions. Outflow boundary conditions are used at the boundary of the computational box (i.e., the planes
$x=-L,+L$, $y=-L,+L$ and $z=-L,+L$).
\item \textbf{Octant:} In octant simulations the simulation box extends from $0$ to $+L$ along the three directions.
We inject SNe in a spherical `star cluster' centred at the origin, and the outcomes are spherically symmetric
in a statistical sense. Therefore, these simulations are statistically equivalent to the full box simulations, but
are computationally less expensive by roughly a factor of $8$. These simulations are only carried out for a large
number of SNe ($N_{\rm OB} \geq 10^{3}$) because of  a larger spatial stochasticity for small number
of SNe; for small $N_{\rm OB}$, an octant may have an effective number of SNe which is substantially different
from $N_{\rm OB}/8$. For precise mass and energy budgeting, we account for the actual mass and energy dumped in by
SNe in all cases. Reflective boundary conditions are used at the faces intersecting within the `star cluster'
(i.e., the planes $x=0$, $y=0$ and $z=0$).
\end{itemize}
\subsection{Supernova Energy Injection}\label{sec:eng-inj}
In our setup, supernovae explode within a `star cluster', a spherical region of radius $r_{\rm cl}$ centred at
the origin of the simulation box. Most young star clusters are $\lesssim 10\ \rm pc$ in size (e.g.,
see \citealt{1999A&AS..139..393L}) but we allow $r_{\rm cl}$ to be larger. A larger $r_{\rm cl}$ crudely
mimics a collection of star clusters that powers global galactic outflows such as in M82 (\citealt{1995ApJ...446L...1O}).
The locations of SNe are chosen randomly, distributed uniformly within a sphere of radius $r_{\rm cl}$, using
the uniform random number generator \texttt{ran2} \citep{1986nras.book.....P}. Supernovae are injected uniformly in
time, with the time separation between successive SNe given by
\begin{equation}
\label{eq:dtSN_tau}
 \delta t_{\rm SN} = \frac{\tau_{\rm OB}}{N_{\rm OB}},
\end{equation}
where $\tau_{\rm OB}$ (chosen to be 30 Myr) is the life time of the OB association and $N_{\rm OB}$ is the
total number of SNe (which equals the total number of O and B stars). \citealt{2010A&A...510A.101F} have shown 
that statistically the supernova rate is uniform. \citealt{1987ApJ...317..190M} also show that a constant mechanical 
luminosity is a good approximation to supernova energy injection. Also, it helps to understand 
the numerical results with simple analytic calculations.

Each SN deposits a mass of $M_{\rm SN} = 5\ \rm M_\odot$ and internal energy
of $E_{\rm SN} =  10^{51}\ \rm erg$ over a sphere of size $r_{\rm SN} = 5\ \rm pc$; the SN energy injection radius is
chosen to prevent artificial cooling losses (see Eq. 7 in \citetalias{2014MNRAS.443.3463S},
corresponding to their thermal explosion model). \citetalias{2014MNRAS.443.3463S}
found that the late time (after a SN enters the Sedov-Taylor stage) results are independent
of whether SN energy is deposited as kinetic or thermal energy (see their Figs. 2 \& 3), so we
simply deposit thermal energy.

Mass and energy injection from each SN is spread in space and time using a Gaussian kernel, such
that the mass and internal energy source terms ($\dot{\rho}_{\rm SN}$ in Eq. \ref{eq.rho} and
$\dot{e}_{\rm SN}$ in Eq.~\ref{eq.energy}) are proportional to
$\exp( - [t-t_{\rm i}]^2/\delta t_{\rm inj}^2)\times \exp( - [\mathbf{x}-\mathbf{x}_{\rm i}]^2/r_{\rm SN}^2)$,
where $i$th SN is centred at $t_{\rm i}$ in time and at
$\mathbf{x}_{\rm i}$ in space. The injection timescale is chosen to be $\delta t_{\rm inj} = \delta t_{\rm SN}/10$.
SN injection with smoothing is found to be numerically more robust and the results are
insensitive to the details of smoothing. In addition to thermal energy, we deposit a subdominant amount
of kinetic energy because the mass that we add in each grid cell (Eq. \ref{eq.rho}) is added at the local velocity.
We account for this additional energy in our energy budget.

We have carried out simulations with different values of initial ambient density ($n_{g0}$), cluster
size ($r_{\rm cl}$) and number of supernovae ($N_{\rm OB}$). The physical size of the simulation box is
chosen according to the number of SNe and the ambient density (based on the adiabatic bubble formula
of \citealt{1987ApJ...317..190M}, the outer shock radius $r_{\rm sb}\propto[N_{\rm OB}t^3/n_{g0}]^{1/5}$).
All our 3-D simulations are listed in Table \ref{table:convergence} (convergence runs for the fiducial parameters,
$N_{\rm OB} = 100$, $n_{g0}=1$ cm$^{-3}$, $r_{\rm cl}=100$ pc) and Table \ref{table:simulations} (all other runs).
\section{Results}
\subsection{The fiducial Run} \label{sec:fiducial}
In this section we describe in detail the morphology and evolution of a superbubble
for number of SNe $N_{\rm OB}=100$, initial gas density $n_{g0} = 1~{\rm cm}^{-3}$, and cluster
radius $r_{\rm cl}=100$ pc, which we choose as our fiducial run.
The assumed parameters are typical of supershells (e.g., \citealt{1979ApJ...229..533H,2014A&A...564A.116S,2011AJ....141...23B}),
but as mentioned earlier, $r_{\rm cl}$ is larger than typical cluster sizes. Our spatial resolution is $\delta L=2.54$ pc
(run R2.5 in Table \ref{table:convergence}). Simulations with different $N_{\rm OB}$ and $n_{g0}$ evolve in a qualitatively similar fashion,
the differences being highlighted in section~\ref{sec:NOB_and_n}.
Numerical resolution quantitively affects our results, although the qualitative trends remain similar. Strict convergence is not
expected because thermal and viscous diffusion are required to resolve the turbulent boundary layers connecting hot
and warm phases (e.g., \citealt{2004ApJ...602L..25K}). A detailed convergence study is presented in Appendix \ref{sec:convergence}.
\begin{figure*}
\includegraphics[width=\linewidth]{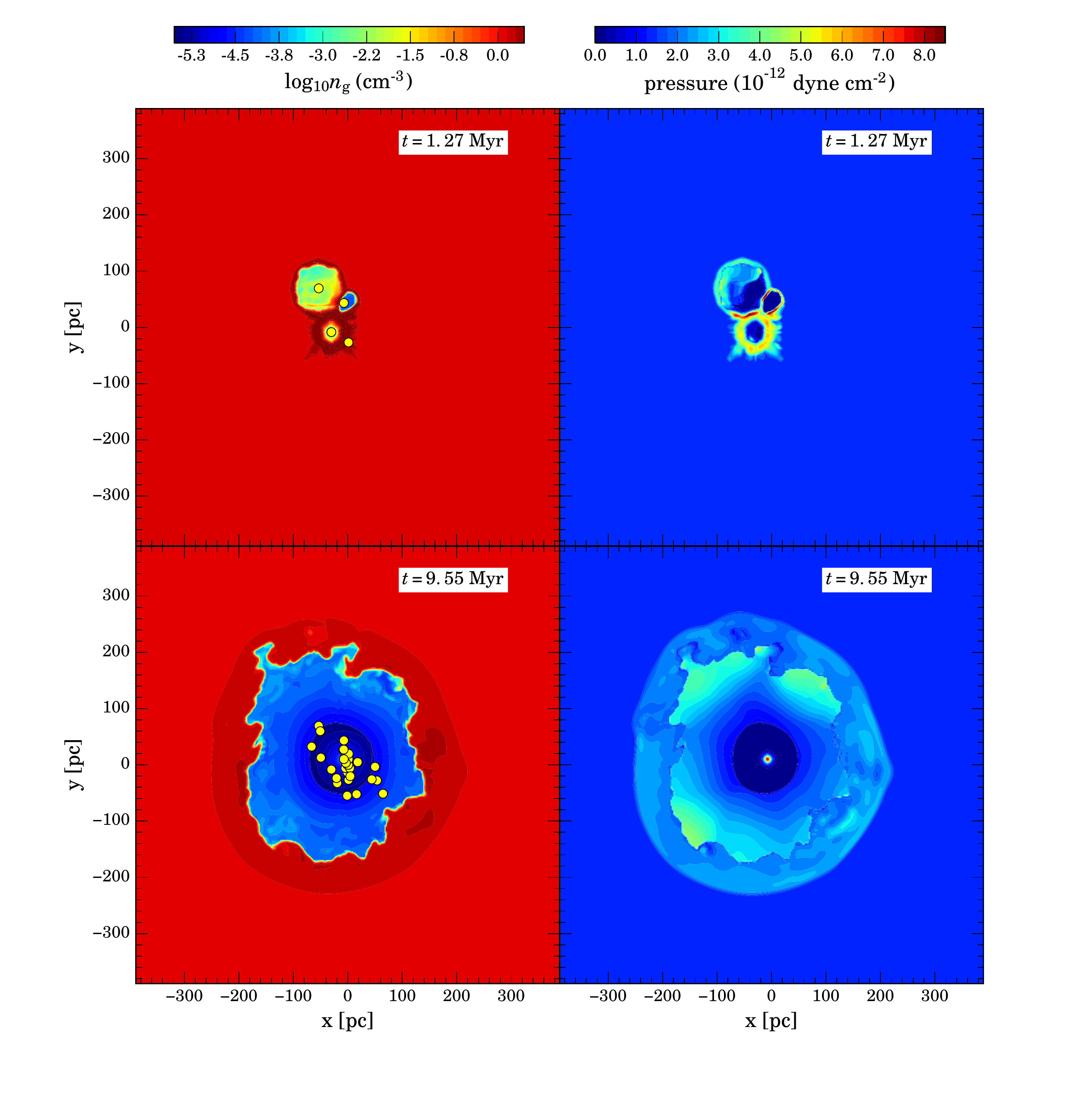}
\caption{Gas density (left panels; $\log_{10} n_g$[cm$^{-3}$]) and pressure (right panels) snapshots in the $z=0$
plane from our fiducial run shown before (top panels) and after (bottom panels) SNe coalesce. The
yellow dots mark the {\it projected} location of SNe in the $z=0$ plane, with four SNe having
exploded by $1.27\ \rm Myr$ and $31$ by $ 9.55\ \rm Myr$. Top panels show that the SNe
at $1.27\ \rm Myr$ are effectively isolated and even at this short time (say, compared to a galaxy's dynamical time)
the pressure within their individual bubbles is smaller than the ISM pressure. The bottom panels show the formation of a superbubble due to the
overlap of several SNe. The pressure inside most of the bubble volume, except at the center, is
larger than the ISM value. Note that a SN has gone off just before $9.55\ \rm Myr$, and it creates
a high pressure sphere right at the center. Also note that while the density scale is logarithmic, the pressure scale is linear.
}\label{fig:np-snapshots}
\end{figure*}
Fig. \ref{fig:np-snapshots} shows the gas density and pressure slices in the midplane of the simulation domain 
at times when SNe are effectively isolated (1.27 Myr) and when they have coalesced (9.55 Myr) to form an 
overpressured superbubble. Since the evolution of a single SN is well known (see, e.g., Figs. 1 \& 2 in 
\citealt{2015ApJ...802...99K}), in order to compare with superbubble evolution we just briefly review the different 
phases of SN evolution. A SN shock starts in the free-expansion phase, moving ballistically till the ejecta sweeps 
up its own mass in the ISM. The next phase is the well-known adiabatic Sedov-Taylor (ST) phase, which transitions to a 
radiative snowplow phase with a thin radiative shell. The radius at which a supernova enters the ST phase can be 
written as
\begin{equation}
r_{\rm ST} \approx 4.3M_{\rm SN,5}^{1/3} n_{\rm g0,1}^{-1/3}\ \rm pc,
\end{equation}
which in all cases is more than twice the grid resolution. Therefore, in our fiducial run we barely resolve the ST phase
of the first few SNe. 
The corresponding ST timescale is 
\begin{equation}
t_{\rm ST} \approx 6\times 10^{-4} M_{\rm SN,5}^{5/6}  E_{\rm SN,51}^{-1/2}  n_{\rm g0,1}^{-1/3}\ \rm Myr.
\end{equation}
For supernovae going off inside a rarified bubble (in which most subsequent SNe explode) $r_{\rm ST}$ is larger and $t_{\rm ST}$ is longer. In the ST phase
the bubble loses pressure adiabatically. The bubble stops expanding by $\sim$0.5 Myr after which the interior pressure falls below the ambient value. In this state
the shock slows down to the sound speed in the ambient medium and becomes a sound wave. The SN fizzles out by $\sim1$ Myr.
The maximum SN bubble size is $\lesssim 50$ pc.

Various stages of a single SN evolution can be seen in the top panels of Fig. \ref{fig:np-snapshots}, which show four
isolated SNe that have exploded by 1.27 Myr. The top-left SN (see the projected
locations of SNe in the top-left panel) is the oldest, followed by the bottom left one; both have faded away, as can be seen from a relatively
high density and low pressure in the bubble region. The other two SNe are younger. The bottom two panels of Fig. \ref{fig:np-snapshots}
show a fully developed superbubble; it is impossible to make out individual SN remnants. Since most stars form in clusters, individual SN remnants
are an exception rather than a norm (e.g., see \citealt{2014IAUS..296..273W}). Most superbubble volume is overpressured (albeit slightly)
relative to the ISM. Thus, superbubbles as a manifestation of overlapping SNe are qualitatively different from isolated SNe.
\subsubsection{Global mass and energy budget}\label{sec:me_budget}
\begin{figure}
\includegraphics[width=1.1\linewidth]{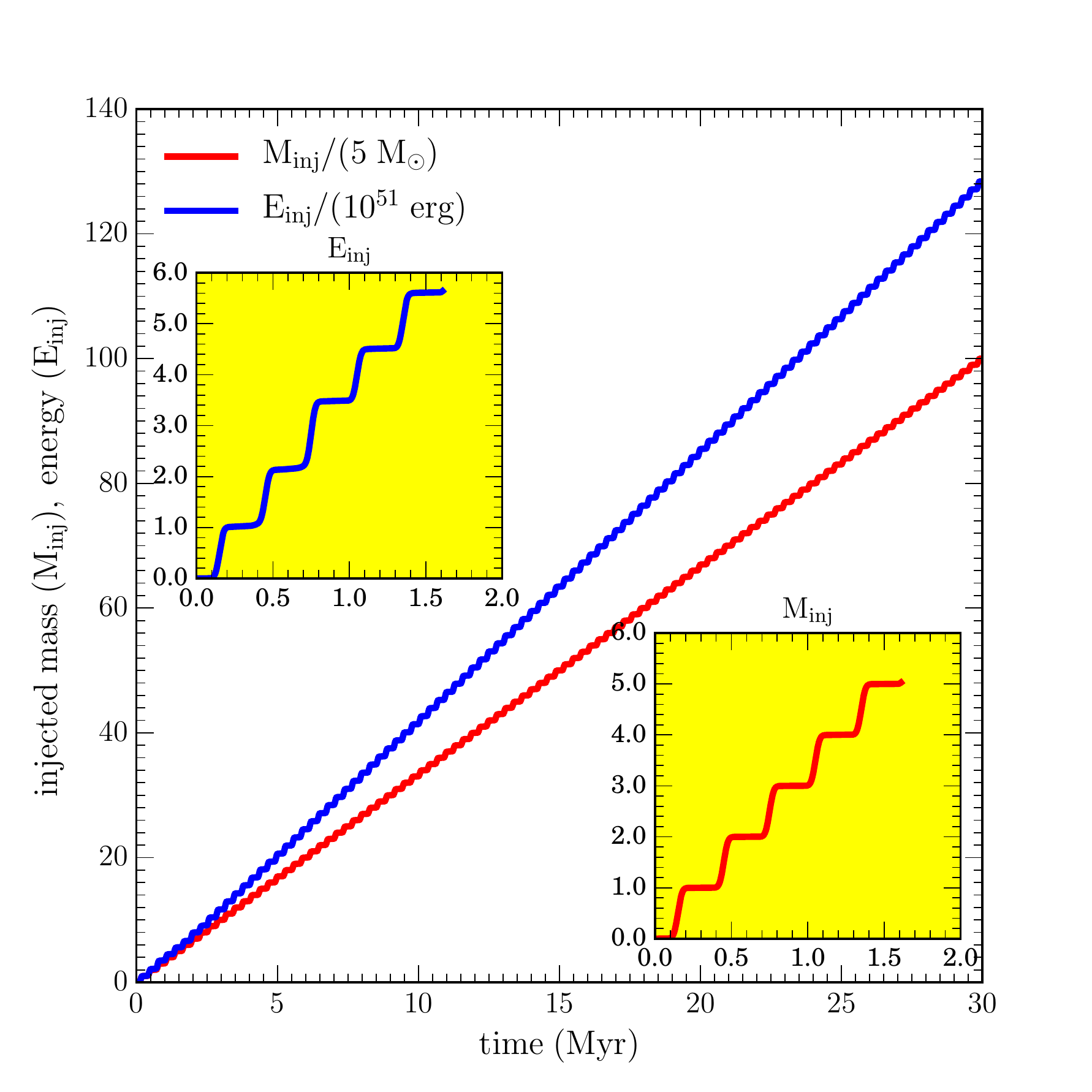}
\caption{Energy (thermal + kinetic) and mass injected in the simulation box (their value at a given time minus the initial value,
normalized appropriately) due to SNe as a function of time for the
fiducial run. Injected mass and energy are normalized ($5\ \rm M_\odot$ for mass and $10^{51}$ erg for energy)
such that every SN adds 1 unit. Total energy injected is larger than just the thermal energy put in
due to SNe by $\approx 30\%$ because kinetic energy is injected in addition to the dominant thermal
energy. The insets at top left and bottom right show a zoom-in of injected energy and mass, respectively.
One can clearly see a unit step in the injected mass and energy for each SN that goes off.
}\label{fig:em-budget}
\end{figure}
A key advantage of using a total energy conserving code like {\tt PLUTO} is that energy is conserved to a
very high accuracy and we can faithfully calculate the (typically small) mechanical efficiency of superbubbles.
Fig. \ref{fig:em-budget} demonstrates that our mass injection (mimicking SNe) adds $100 M_{\rm SN}$ by 30 Myr, the
intended amount. The energy added is higher by $\approx 30\%$ because, as mentioned earlier, the mass added
by the density source term (Eq. \ref{eq.rho}) is added at the local velocity, and hence mass addition leads to
the addition of kinetic energy.
\begin{figure}
\includegraphics[width=1.05\linewidth]{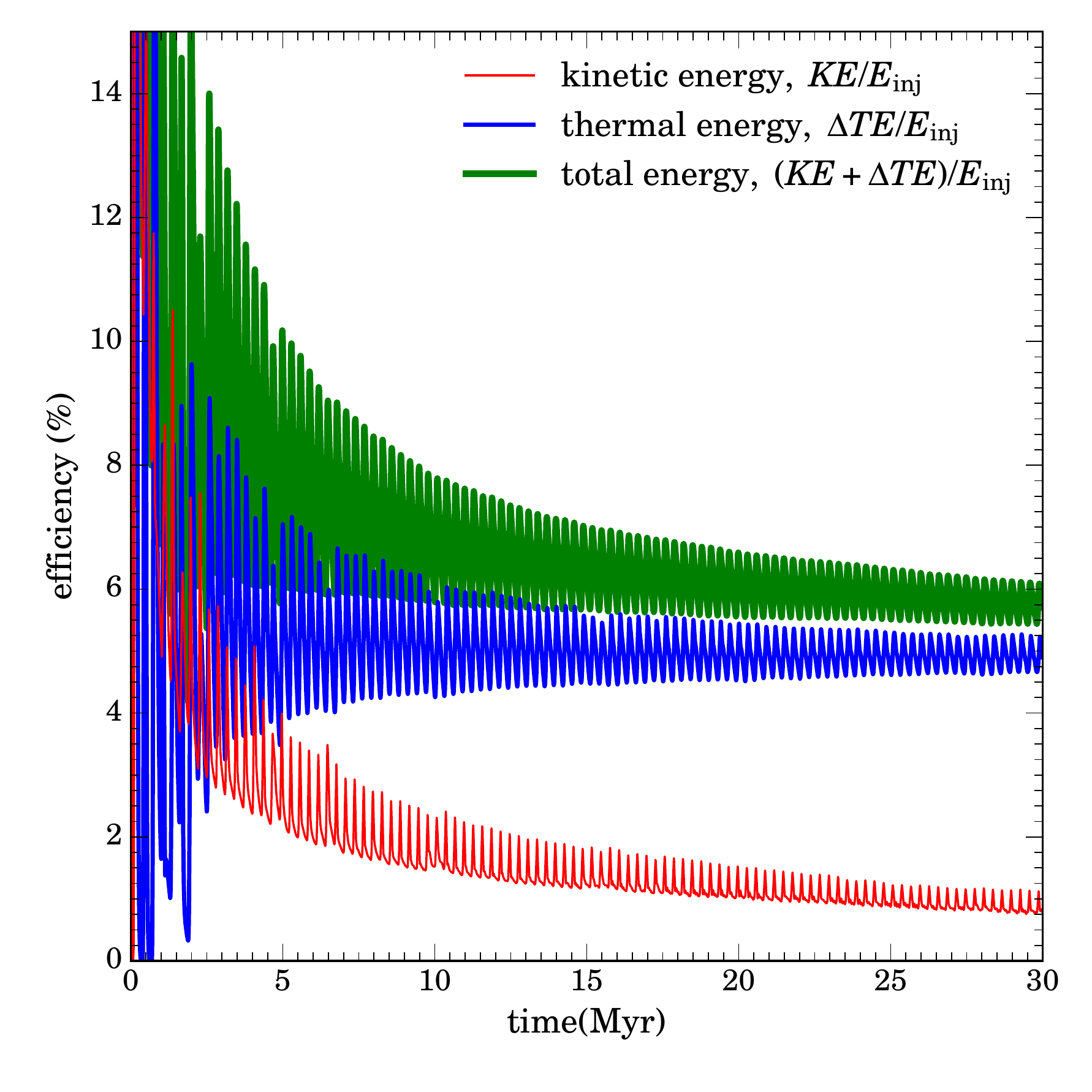}
\caption{The fraction (percentage) of injected energy retained as kinetic energy and thermal energy
of gas inside the simulation box.  At the end of the simulation the gas retains a small fraction, $\approx 1\%$
and $\approx 5\%$ of the total injected energy as kinetic and thermal energy respectively. The periodic spikes in
energies correspond to individual SNe going off. In the legend, $KE$ stands for the kinetic energy and $\Delta TE$
for the change in thermal energy within the computational domain.
}\label{fig:eff-fiducial}
\end{figure}
Fig. \ref{fig:eff-fiducial} shows thermal, kinetic, and total energy efficiency as a function of time for the fiducial run.
Energy efficiency is defined as the ratio of excess energy (current minus initial) in the simulation domain and the
total energy injected by SNe. The energy efficiency that is higher at early times, decreases and asymptotes
to a small value. Due to efficient
cooling, most ($\approx 95\%$ by 30 Myr) of the deposited energy is lost radiatively. Out of the
remaining 5,  $ \approx 4\%$ is retained as the thermal energy and
$1\%$ is retained as the kinetic energy of the gas. In terms of the energy
deposited by a single supernova, the total (kinetic+thermal) energy retained is
$\approx 6\ E_{\rm SN}$.
\subsubsection{Density-pressure phase diagram}
A bubble (associated with both an individual SN and a superbubble) remains hot and dilute for a long time
(several Myr) but is not overpressured with respect to the ISM for a similar duration.
The strength of the bubble pressure compared to that of the ambient medium is a good indicator of its strength.
As pressure decreases with the expansion of the bubble, it will no longer be able to sustain a strong forward shock
and will eventually degenerate into a sound wave.
\begin{figure}
\includegraphics[width=1.1\linewidth]{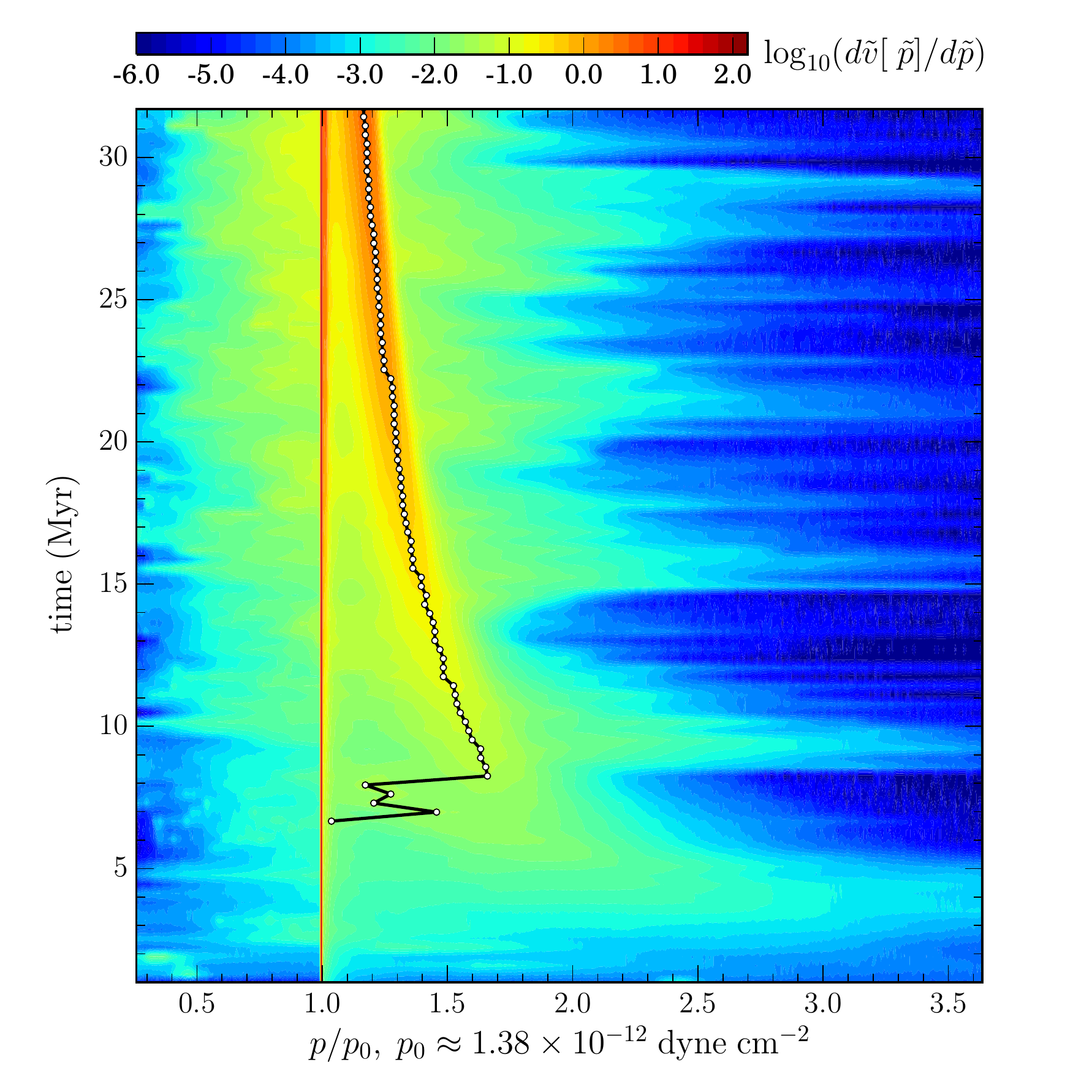}
\caption{Volume distribution of pressure (along horizontal axis; normalized to the initial value $p_0$) at different times
(along vertical axis) for the fiducial run. Color represents the volume fraction
($\log_{10} (d\tilde{\varv}[\tilde{p}]/d\tilde{p})$; $\tilde{\varv}=\varv/V$, where $V=8L^3$ is the volume of the simulation box; $\tilde{p}=p/p_0$;
bin-size in pressure $\Delta \tilde{p}=0.007$) of different pressures at all times. The vertical red line at unity
corresponds to the large volume occupied by the ambient unperturbed ISM. The circles
connected by a solid line mark the location of the local pressure maximum on the higher side.
Before $5\ \rm Myr$ a coherent overlap of isolated SNe has not happened and a distinct structure
in the pressure distribution does not appear.}
\label{fig:pressure-histogram}
\end{figure}
Fig. \ref{fig:pressure-histogram} shows the volume distribution of pressure at all times for the fiducial run.
At $t=0$ all the gas is at the ambient ISM pressure (indicated by the vertical red line at $1.38 \times 10^{-12}$
dyne cm$^{-2}$). Because of a very short-lived high pressure (Sedov-Taylor)
phase and a small volume occupied by the very overpressured gas, the volume fraction of gas with
pressure $\gtrsim 5\times 10^{-12}$ is small at all times. Before few Myr there is no coherent (in time) structure in
the pressure distribution. After overlap of SNe and the formation of a superbubble, there is a  coherent
high pressure peak (shown by the solid line marked by small circles) in the volume distribution that decreases
with time. Bubble pressure decreases because of radiative and adiabatic losses. As the input energy is spread
over a larger and larger volume the bubble pressure decreases and eventually (much later than for an isolated supernova)
becomes comparable to the ambient pressure. At this stage the shell propagates as a sound wave. In short, the first 
few SNe behave as if they are isolated, and as their remnants grow in size they start overlapping and create a superbubble.
In Fig.~\ref{fig:pressure-histogram}, till $4-5$ Myr, the ambient ISM is the most dominant phase. The overlapping 
of SNe leads to the formation of a second dominant branch in the pressure plot, which is at a higher value than the 
ambient pressure.
\begin{figure*}
\includegraphics[width = 1.0\linewidth]{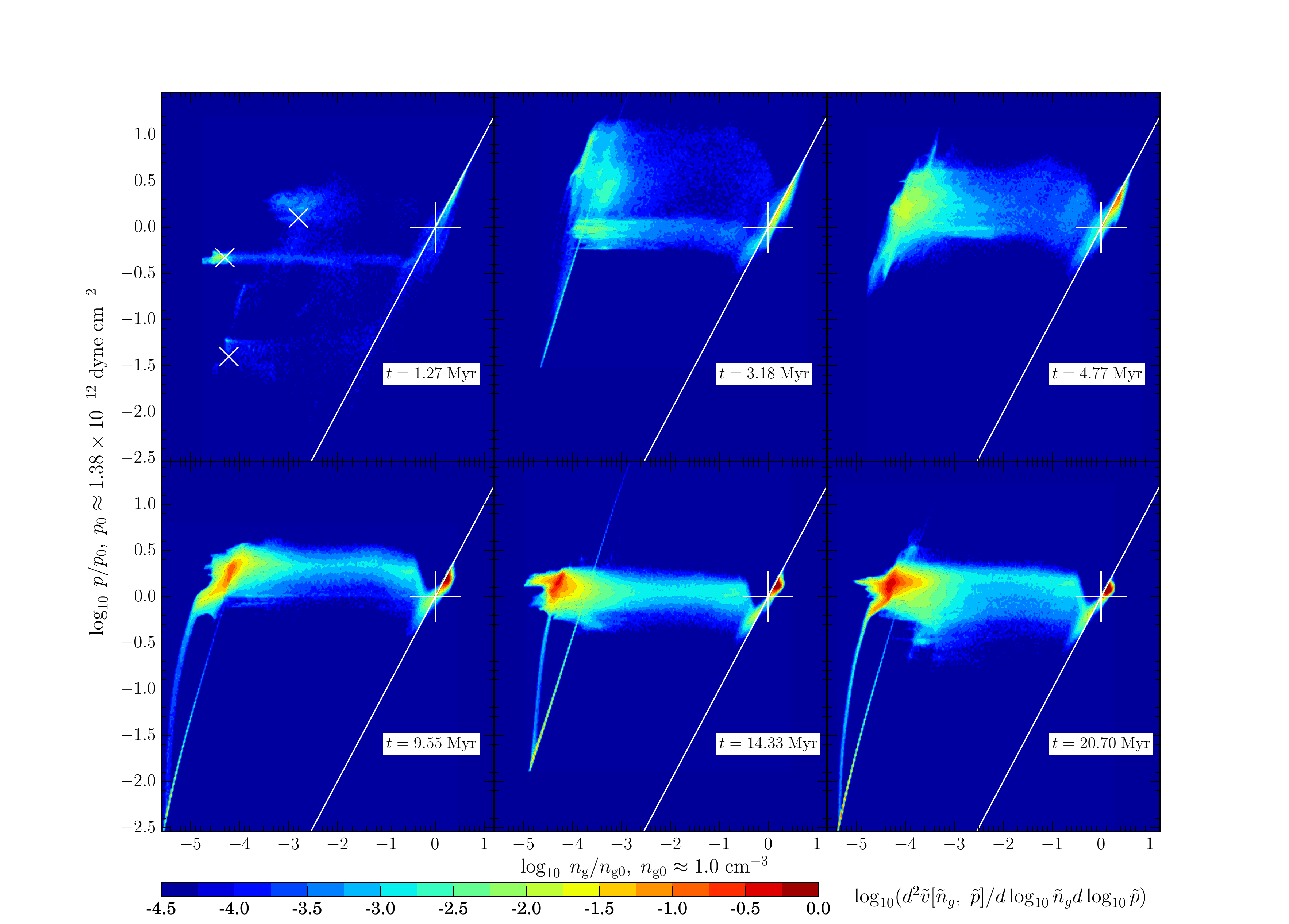}
\caption{Volume distribution of pressure and gas number density ($d^2 \tilde{\varv}/d\log_{10}\tilde{n}_g d\log_{10}\tilde{p}$; $\tilde{\varv}=\varv/V$, $
\tilde{n}_g = n_g/n_{g0}$, $\tilde{p}=p/p_0$;
the bin size $\Delta \log_{10} \tilde{n}_g=0.03$, $\Delta \log_{10} \tilde{p}=0.01$) for the fiducial run at different times. At late times there are two peaks
in the distribution function corresponding to the WNM (ambient ISM at $10^4$ K) and the hot bubble
(at $\sim 10^6-10^8$ K). At 1.27 Myr we can see the signatures of non-overlapping SNe fizzling out.
Later, after about 5 Myr, we see the formation of a low density and (slightly) over pressured
superbubble. At some times we see streaks with $p \propto n_g^{5/3}$, representing adiabatic
cooling of recent SNe ejecta expanding in the low density cavity. The white line
shows a temperature of $10^4$ K, `+' represents the ambient pressure and density, and `x's at 1.27 Myr represent
bubbles corresponding to individual SNe (the bottom-right SN in the top-left panel of Fig. \ref{fig:np-snapshots} is very
young and not clearly seen).
}\label{fig:pv-slice-collage}
\end{figure*}

Fig.~\ref{fig:pv-slice-collage} shows the evolution of gas in the pressure-density
plane. The white plus (+) at $n_g/n_{g0},~p_g/p_0=1$ marks the location of the ambient ISM.
At early times SNe are isolated as evident from the multiple bright areas (marked with white `x')
in the $p-n_g$ distribution at $t = 1.27 \ \rm Myr$.
Significant volume is occupied by gas at the ambient temperature ($10^4$ K), which represents dense/isothermal radiative shells of isolated SNe
at early times and weak outer shock at later times. As the entire cluster
volume is filled with hot gas, it forms an extended hot bubble, and the $p-n_g$ plane shows a bimodal
volume distribution in which the shell/ISM gas is on right and the hot ($\sim 10^8$ K) and rarified gas in the bubble is on left.
As the superbubble ages, the hot ($\sim 10^8$ K) and warm ($\sim 10^4$ K) phases reach rough pressure equilibrium (most of the
superbubble volume is still slightly overpressured; see Fig. \ref{fig:pressure-histogram}). However, the bubble gas density,
even at late times, is $\sim 4$ orders of magnitude smaller relative to the ambient ISM value. In some snapshots (bottom three panels)
we see a straight line with slope equal to $\gamma=5/3$ stretching from low density/pressure to the peak in the hot gas distribution. These streaks
represent adiabatic hot winds launched by continuous (for a short time $\delta t_{\rm inj}$) SN energy injection (see section \ref{sec:eng-inj})
inside the dilute hot bubble (see the very low density/pressure sphere at the center in the bottom panels of Fig. \ref{fig:np-snapshots}). The curved streak
at low pressure/density is due to smaller energy injection at the beginning (and end) of SN energy injection (recall that energy injection
follows Gaussian smoothing in time; see section \ref{sec:eng-inj}). These streaks are an
artefact of our smooth SN energy injection.
\subsubsection{Average profiles \& overpressure-fraction}\label{sec.radius}
\begin{figure}
\includegraphics[width = 1.1\linewidth]{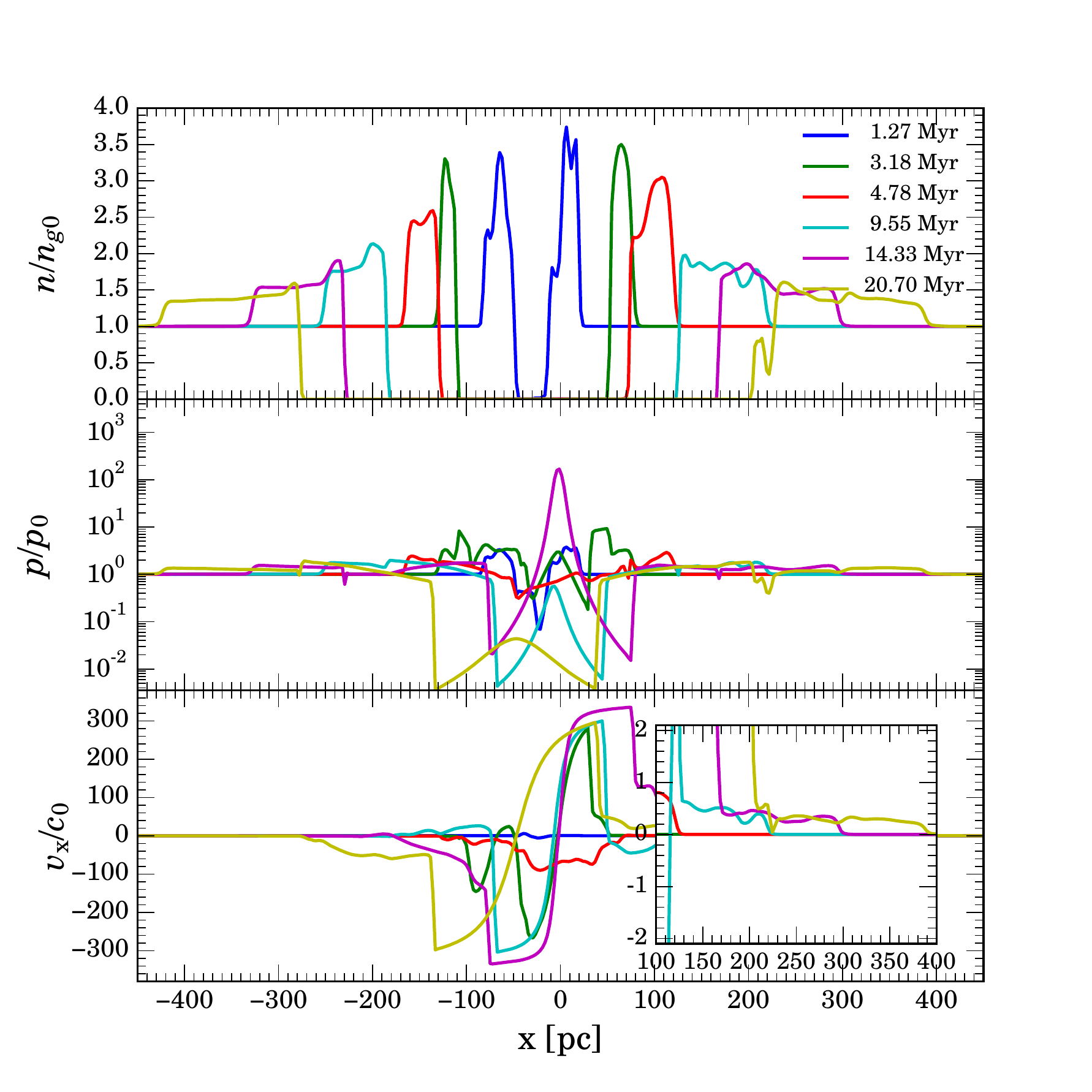}
\caption{Gas density, pressure and $x-$ velocity profiles along the $x-$axis ($y=z=0$) for the fiducial run at various times. The swept-up shell
density decreases with time as the superbubble weakens and eventually the shell propagates at the sound speed in the ambient medium
($c_0 \approx 15$ km s$^{-1}$). As seen in Fig. \ref{fig:np-snapshots}, the bubble density is $\sim 4$ orders of magnitude smaller than the ambient
value. The main bubble pressure decreases with time, except during SN injection, during which a high pressure core and an adiabatic
wind with large velocity and small pressure (similar to \citealt{1985Natur.317...44C}) forms (the streaks seen in some panels of Fig.
\ref{fig:pv-slice-collage} are also a signature of this). The inset in the lowest panel shows that the dense shell propagates at about half the sound
speed in the ambient ISM, but the velocities in the low density bubble are much higher.
}\label{fig:1D-profiles}
\end{figure}
The radius evolution of a single SN remnant inside a uniform medium is well known. The radius
expands differently with time in each of the free-expansion, Sedov-Taylor, pressure-driven snowplow and
momentum-conserving phases (e.g., \citealt{1972ApJ...178..159C}). The radius evolution of a superbubble
is qualitatively different from the radius evolution of a single SN because of the continuous
injection of mechanical energy till the lifetime of the OB association (e.g., \citetalias{2014MNRAS.443.3463S}).
A large bubble pressure is maintained until the energy (only a small fraction of it is retained due to radiative losses)
is spread over a large volume.

The bubble retains only a fraction of injected mechanical energy because of radiative losses. For simplicity, 
the effects of radiative losses can be incorporated in the adiabatic relations using a mechanical 
efficiency factor, $\eta_{\rm mech}$. The superbubble radius ($r_{\rm sb}$) and velocity ($\varv_{\rm sb}=dr_{\rm sb}/dt$)
evolves with time as (Eq. 5 of \citealt{1977ApJ...218..377W})
\begin{eqnarray}
\label{eq:r_sb}
&&r_{\rm sb} \approx  58~{\rm pc}~  \eta^{1/5}_{{\rm mech},-1}  \left ( \frac{  E_{\rm SN,51}
N_{\rm OB,2}  }{ \tau_{\rm OB,30}n_{g0}}  \right )^{1/5}  t_{\rm Myr}^{3/5} , \\
\label{eq:v_sb}
&& \varv_{\rm sb} \approx 34~{\rm km~s}^{-1}~ \eta^{1/5}_{{\rm mech},-1} \left ( \frac{ E_{\rm SN,51}
N_{\rm OB,2} }{ \tau_{\rm OB,30} n_{g0} } \right )^{1/5} t_{\rm Myr}^{-2/5},
\end{eqnarray}
where $E_{\rm SN,51}$ is the SN energy scaled to $10^{51}$ erg, $N_{\rm OB,2}$ is the number of OB stars
in units of 100, $\tau_{\rm OB,30}$ is the age of OB association in units of 30 Myr, $n_{g0}$ is  the ambient gas density
in cm$^{-3}$, and $t_{\rm Myr}$ is time in Myr. The mechanical energy retention efficiency $\eta_{\rm mech,-1}$ is
scaled to 0.1. The supershell velocity can be expressed in terms of its radius as
\begin{equation}
\label{eq:v_r_sb}
\varv_{\rm sb} \approx 34~{\rm km~s}^{-1}~ \left( \frac{ \eta_{{\rm mech},-1} E_{\rm SN,51}
N_{\rm OB,2}} {\tau_{\rm OB,30} n_{\rm g0} r_{\rm sb,58 pc}^2 } \right)^{1/3}.
\end{equation}
The superbubble weakens after the the outer shock speed becomes comparable to the sound speed; i.e., $\varv_{\rm sb} \approx c_0$
($c_0 \equiv [\gamma k_BT_0/\mu m_p]^{1/2}$ is the sound speed in the ambient ISM). Thus, using Eq. \ref{eq:v_sb}, the fizzle-out time is
\begin{equation}
\label{eq:tiff}
t_{\rm fiz} \approx 21.3~{\rm Myr}~\eta^{1/2}_{{\rm mech},-1} \left ( \frac{ E_{\rm SN,51}
N_{\rm OB,2} }{ \tau_{\rm OB,30} n_{g0} } \right )^{1/2} c_{0,1}^{-5/2},
\end{equation}
where $c_{0,1}$ is the sound speed in the ambient medium in units of 10 km s$^{-1}$.
\begin{figure}
\includegraphics[width = 1.0\linewidth]{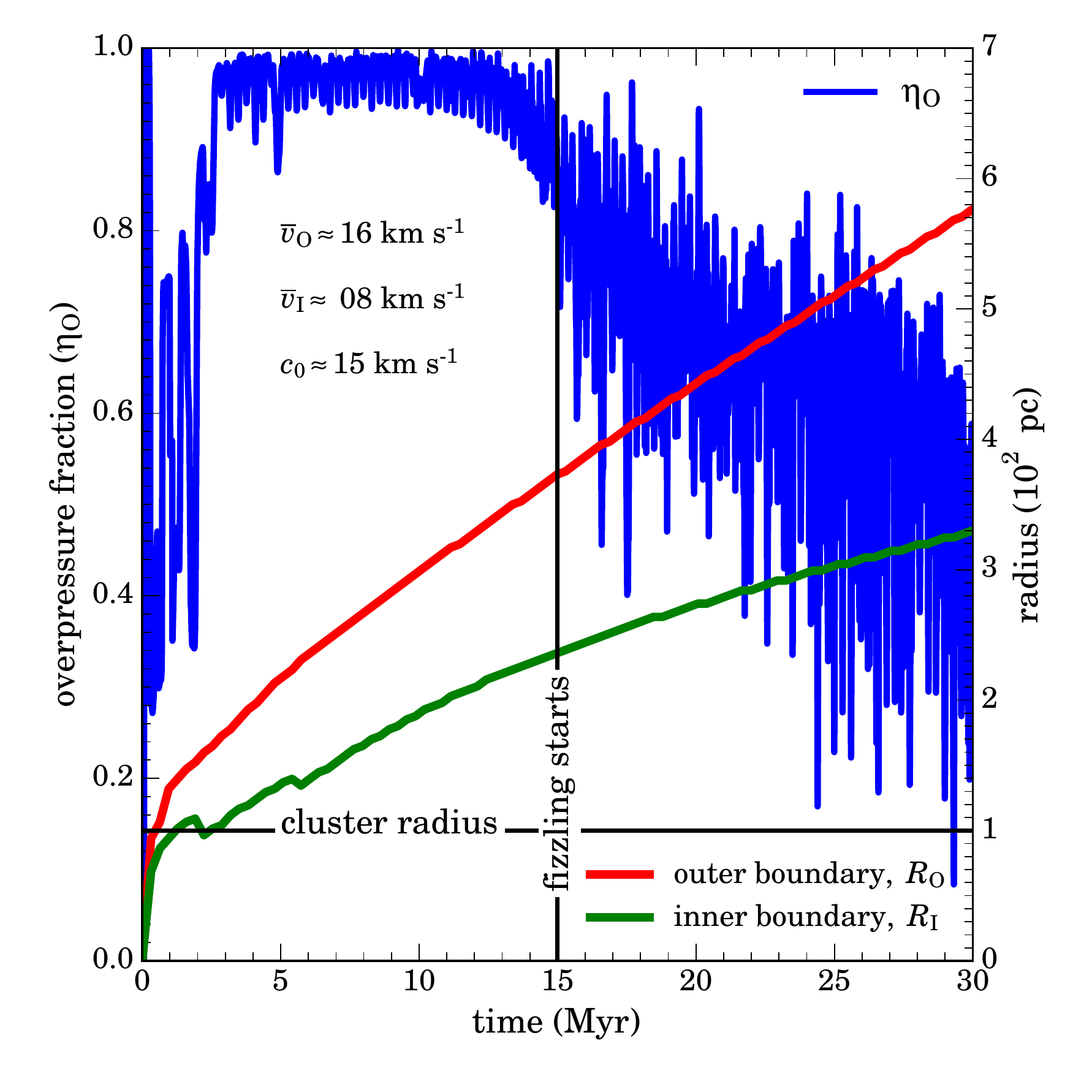}
\caption{The inner (green line) and outer (red line) radius of the superbubble shell as a function of
time for the fiducial run. The blue line shows the overpressure fraction ($\eta_O$) as a function of time.
The superbubble starts to fizzle out when the overpressure fraction starts falling from $\approx 1$, which happens
around 15 Myr. The average outer shell velocity is comparable to the ISM sound speed; the inner shell speed is smaller.
The bottom panel of Fig. \ref{fig:1D-profiles} shows that at late times the shell material moves at $\approx c_0/2$,
similar to the inner shell speed. The outer shell velocity is higher, $\approx c_0$, consistent with the
shell density decreasing in time.}
\label{fig:rvst_n_opf}
\end{figure}
Fig. \ref{fig:1D-profiles} shows density, pressure and $x-$ velocity profiles along the $x-$ axis for the fiducial run at the same times as the panels
in Fig. \ref{fig:pv-slice-collage}.
The evolution of various profiles is as expected. The shell become weaker  and slower with time and eventually propagates at roughly
the sound speed of the
ambient medium ($c_0$). Time evolution of the angle-averaged (unlike Fig. \ref{fig:1D-profiles}, which shows a cut along $x-$ axis)
inner and outer shell radii (see Appendix \ref{app:rad}) is shown in Fig. \ref{fig:rvst_n_opf}.

The radius and velocity evolution of bubbles is critically dependent on the presence of radiative
losses (encapsulated by $\eta_{\rm mech}$; \citealt{1977ApJ...218..377W}). In order to assess the
strength of a superbubble, it is useful to define an overpressure volume fraction ($\eta_{\rm O}$) as
\begin{equation}\label{eq.volfrac}
\eta_{\rm O} = \frac{V_{>}}{V_{>}+V_{<}},
\end{equation}
where $V_{>}$ is the volume occupied by gas at pressure $p > 1.5p_0$ and $V_{<}$ is the volume occupied
by gas at $p<p_0/1.5$ ($p_0$ is the ambient ISM pressure; the choice of 1.5 is somewhat arbitrary).
Thus $\eta_{\rm O}$ gives the fraction of volume occupied by high pressure, hot and dilute bubble gas. Since
we exclude gas close to the ambient ISM pressure in its definition, $\eta_{\rm O}$ is independent of the
computational domain and characterizes the bubble pressure. In Fig.~\ref{fig:rvst_n_opf} we also show the
evolution of the hot volume fraction ($\eta_{\rm O}$) as a function of time for the fiducial run. The hot volume fraction
drops initially when SNe have not overlapped, but reaches
unity after $\approx 3$ Myr, and starts decreasing rapidly after radiative losses become significant and the bubble pressure
become comparable to the ISM pressure (or equivalently, the shock velocity becomes comparable to the ISM sound speed).
The nature of the hot volume fraction evolution is discussed in more detail in sections \ref{sec:rcl} and \ref{sec:eta_nob_ng}.
\begin{table*}
\caption{Parameters of our 3-D simulations}
\begin{centering}
 \begin{tabular}{c c c c c c c c c c c}
 \hline
$N_{\rm OB}$ &$n_{\rm g0}$ &$r_{\rm cl}$ &$L$ &$N$ &$\delta L$ & Sim Type &$\delta t_{\rm SN}$ &$E_{\rm inj}^\ddag$ &$\eta_{\rm mech}^\dag$ &$\eta_{\rm O}^\dag$\\
      &($\rm cm^{-3}$) &($\times 10^{2}\ \rm pc$) &(pc) & &(pc) & F/O &(Myr) &($E_{SN,51})$& ($\%$) & \\
\hline
100 & 0.1 & 1.0 & 1038 & 410 &  2.54 &   O & 0.24 & 12 & 12.66 &  1.00 \\
100 & 0.3 & 1.0 & 876 & 700 &  2.50 &   F &  0.24 & 137 & 11.37 &  0.98 \\
100 & 0.5 & 1.0 & 779 & 620 &  2.51 &   F &  0.24 & 137 &  8.70 &  0.69 \\
100 & 0.8 & 1.0 & 714 & 570 &  2.51 &  F &  0.24 & 136 &  6.74 &  0.61 \\
100 & 2.0 & 1.0 & 698 & 570 &  2.45 &  F &  0.24 & 135 &  3.28 &  0.22 \\
100 & 1.0 & 0.0 & 714 & 580 &  2.46 &   F &  0.24 & 123 &  4.66 &  0.37 \\
100 & 1.0 & 0.7 & 649 & 300 &  4.33 &   F &  0.24 & 131 &  4.49 &  0.41 \\
100$^\star$ & 1.0 & 1.0 & 649 & 512 &  2.54  & F &  0.24 & 136 &  5.72 &  0.47 \\
100 & 1.0 & 1.5 & 649 & 300 &  4.33 &   F &  0.24 & 130 &  3.93 &  0.36 \\
100 & 1.0 & 0.3 & 601 & 512 &  2.35 &   F &  0.24 & 138 &  6.38 &  0.44 \\
\hline
1000 & 1.0 & 1.0 & 1136 & 450 &  2.54  & O &  0.024 & 134 &  5.34 &  0.94 \\
1000 & 2.0 & 1.0 & 1055 & 420 &  2.51   & O &  0.024 & 134 &  3.17 &  0.44 \\
1000 & 3.0 & 1.0 & 974 & 385 &  2.54  & O &  0.024 & 134 &  1.96 &  0.21 \\
1000 & 4.0 & 1.0 & 909 & 360 &  2.54  & O &  0.024 & 134 &  1.25 &  0.14 \\
1000 & 5.0 & 1.0 & 876 & 350 &  2.50  & O &  0.024 & 133 &  0.83 &  0.12 \\

\hline
$10^4$ & 5.0 & 1.0 & 1006 & 400 &  2.52  & O &  0.0024 & 1408 &  1.80 &  0.86 \\
$10^4$ & 6.0 & 1.0 & 1006 & 400 &  2.52   & O &  0.0024 & 1406 &  1.44 &  0.58 \\
$10^4$ & 8.0 & 1.0 & 974 & 385 &  2.54 &  O &  0.0024 & 1405 &  1.07 &  0.16  \\
$10^4$ & 9.0 & 1.0 & 974 & 385 &  2.54 &  O &  0.0024 & 1398 &  0.38 &  0.08  \\
$10^4$ & 10.0 & 1.0 & 860 & 340 &  2.54   & O &  0.0024 & 1404 &  0.68 &  0.07 \\
$10^4$ & 10.0 & 0.5 & 649 & 512 &  2.54   & F &  0.0024 & 16274 &  1.45 &  0.44 \\
$10^4$ & 10.0 & 1.0 & 649 & 512 &  2.54   & F &  0.0024 & 16385 &  1.49 &  0.48 \\
$10^4$ & 10.0 & 1.5 & 649 & 512 &  2.54   & F &  0.0024 & 15151 &  1.48 &  0.47 \\
$10^4$ & 10.0 & 2.0 & 649 & 512 &  2.54   & F &  0.0024 & 13971 &  1.38 &  0.45 \\
$10^4$ & 10.0 & 2.5 & 649 & 512 &  2.54   & F &  0.0024 & 13173 &  1.29 &  0.43 \\
$10^4$ & 10.0 & 3.0 & 649 & 512 &  2.54   & F &  0.0024 & 12409 &  1.23 &  0.44 \\

\hline
$10^5$ & 1.0 & 1.0 & 2110 & 850 &  2.48 &   O &  $2.4 \times 10^{-4}$ & 16036 &  4.58 &  0.92 \\
$10^5$ & 10.0 & 1.0 & 1363 & 550 &  2.48  & O &  $2.4 \times 10^{-4}$ & 15955 &  1.65 &  0.92 \\
$10^5$ & 20.0 & 1.0 & 1233 & 490 &  2.52  & O &  $2.4 \times 10^{-4}$ & 15899 &  1.06 &  0.83 \\
$10^5$ & 30.0 & 1.0 & 1136 & 450 &  2.54  & O &  $2.4 \times 10^{-4}$ & 15843 &  0.78 &  0.47 \\
$10^5$ & 40.0 & 1.0 & 1071 & 425 &  2.52   & O &  $2.4 \times 10^{-4}$ & 15786 &  0.57 &  0.09 \\
$10^5$ & 50.0 & 1.0 & 1006 & 400 &  2.52   & O &  $2.4 \times 10^{-4}$ & 15732 &  0.42 &  0.05 \\
\hline
\end{tabular}
\end{centering}
\begin{flushleft}
{\scriptsize $^\ddag$ The actual energy injected in units of $10^{51}$ erg; this can be slightly different from
$N_{\rm OB} E_{\rm SN}$ because of extra kinetic energy injection; octant runs should inject $\approx E_{\rm SN}/8$
as only one octant is simulated.\\ $^\dag$ $\eta_{\rm mech}$ (Eq. \ref{eq:mech_eff}) and $\eta_{\rm O}$ (Eq. \ref{eq.volfrac})
are averaged over $t=29-30\ \rm Myr$.\\ $^\star$ The fiducial run.\\}
\end{flushleft}
\label{table:simulations}
\end{table*}

\subsection{Effects of thermal conduction}
\begin{figure*}
\includegraphics[width = \linewidth]{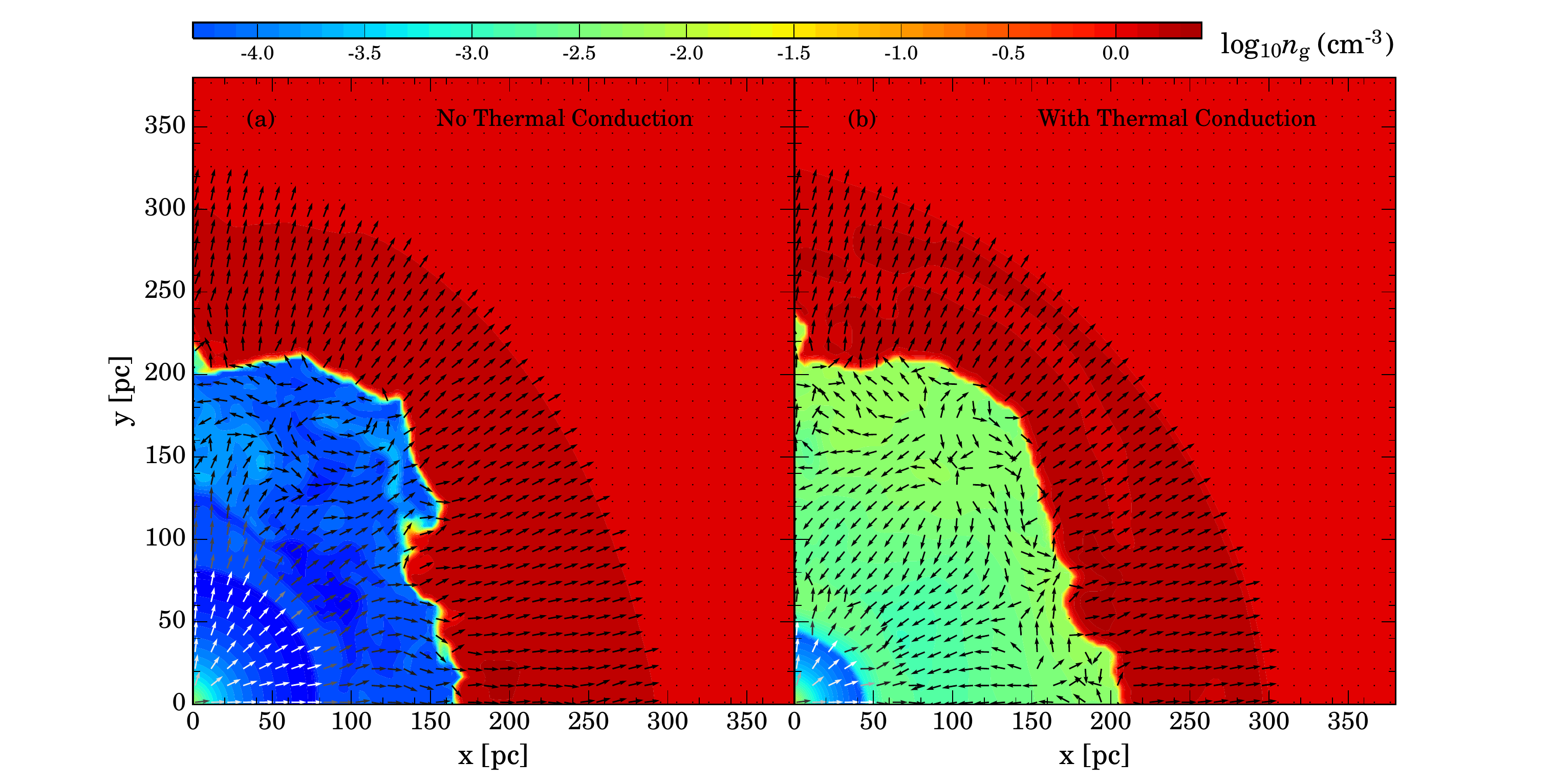}
\caption{A density contour plot and a quiver plot showing the projection of the velocity unit vector ($\hat{\varv}$) in the x-y plane for the 
fiducial simulation with (right panel) and  without (left panel) conduction. Note that in the simulation with conduction there
is a region $\approx 100-150\ \rm pc$ in which the flow is directed inwards, whereas such a flow is absent in the run without
conduction; this flow occurs as conduction leads to evaporation of material from the shell into the bubble. The parameters of these runs 
are: number of SNe
$N_{\rm OB}=100$, initial gas density $n_{g0} = 1~{\rm cm}^{-3}$, and cluster radius $r_{\rm cl}=100$ pc. The snapshots are
at $\approx 14.33\ \rm Myr$.}\label{fig:conduction-flow}
\end{figure*}

We have done the fiducial run with the isotropic thermal conduction module in \texttt{PLUTO} code, which implements Spitzer 
and saturated thermal conduction based on super time stepping (STS, \citealt{alexiades1996super}; $\nu_{\rm STS}=0.01$). 
Matter evaporates from the cold shell to the interior of the hot bubble (made of shocked SNe) due to thermal conduction, 
as shown analytically by \citealt{1975ApJ...200L.107C} (see also the right panel of Fig. 9 in \citetalias{2014MNRAS.443.3463S}). 
Fig.~\ref{fig:conduction-flow} shows the density snapshots and projected velocity unit vectors for the fiducial run with (right panel) and without 
(left panel) conduction. The density in the hot bubble is much higher with conduction due to the evaporative flow from the dense shell 
to the hot bubble, as indicated by the velocity unit vectors in the right panel. Such a flow is absent in the run without conduction.
The maximum temperature
reached by the gas with conduction is much smaller than without it (c.f. Fig.~\ref{fig:temp_luminosity}). 
Overall, we find that thermal conduction does not affect the dynamics of the shell (e.g., its radius and velocity) but affects the temperature 
distribution of gas within the shell, which can influence its emission/absorption signatures. Since superbubble dynamics is unaffected by thermal 
conduction, we do not include it in the rest of our simulations.

\subsection{Comparison with 1-D simulations} \label{sec:1D}
Most of supernova and superbubble studies are carried out in spherical 1-D geometry because these systems are spherical
(although only crudely) and very high resolution runs can be done. We want to compare our more realistic
3-D simulations (albeit with much lower resolution compared to the modern 1-D simulations) with 1-D runs
to highlight the similarities and differences between the two.

For a realistic comparison of 1-D spherical and 3-D Cartesian runs, we run a 3-D simulation in which we 
explode all SNe at the origin (i.e., $r_{\rm cl}=0$). Both the 1-D and 3-D runs have the same resolution as the
fiducial run ($\delta L=\delta r$=2.54 pc; the only difference between this 3-D simulation and the fiducial run
is that here $r_{\rm cl}=0$). As discussed in section \ref{sec:me_budget}, the amount of total mechanical energy
injected in the box is slightly larger than $N_{\rm OB} \times E_{\rm SN}$ because of extra kinetic energy that we
put in due to mass addition at the local velocity. For an exact comparison of our 1-D and 3-D runs we match the
total energy injected in our 1-D and 3-D runs (by slightly scaling $E_{\rm SN}$ for the 1-D run).
\begin{figure*}
\includegraphics[width = \linewidth]{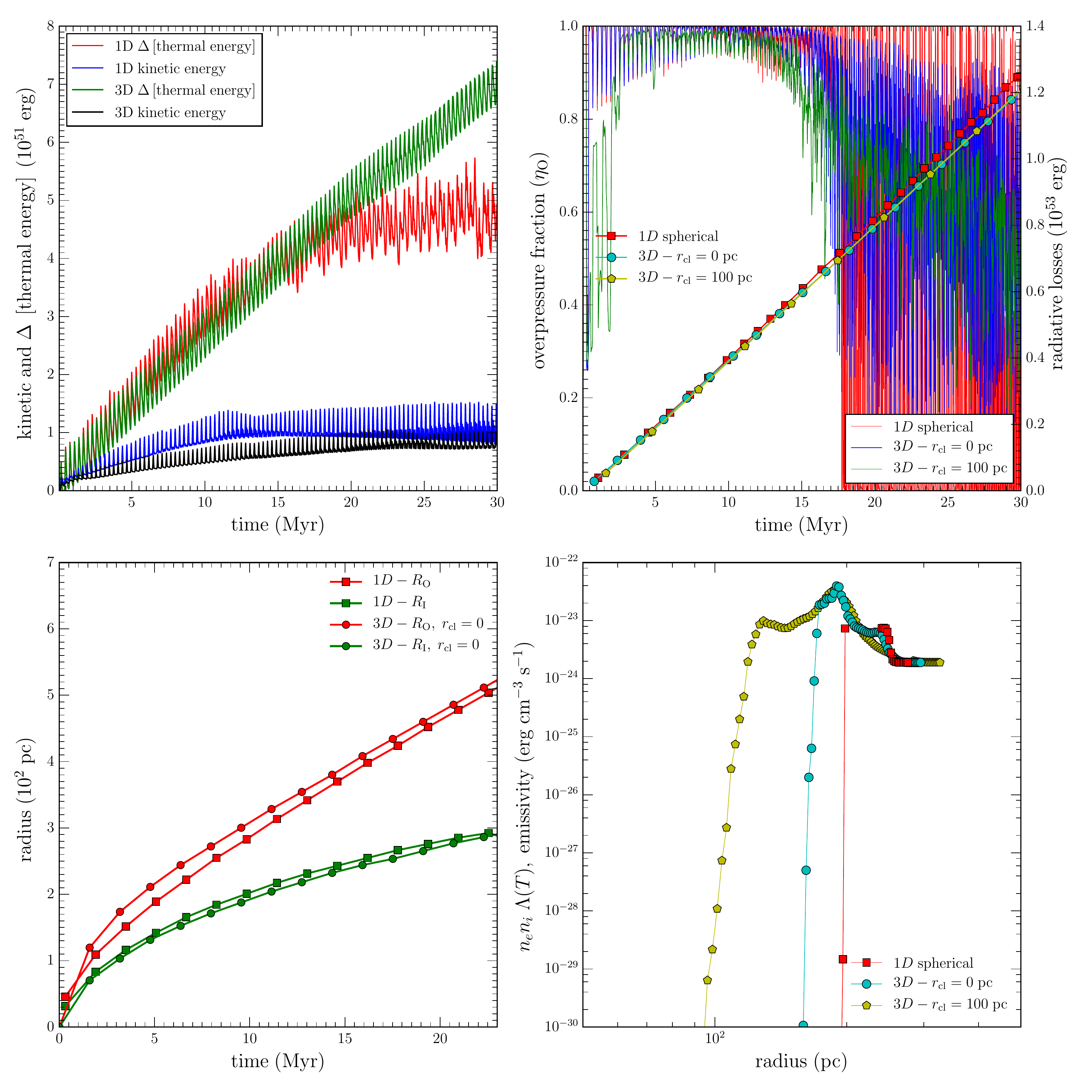}
\caption{A comparison of 3-D (with $r_{\rm cl}=0$ pc; $r_{\rm cl}=100$ pc run is also shown for the right panels)
and 1-D spherical simulations.
The top-left panel shows the kinetic and thermal energy added to the box by SNe.
The top-right panel shows the overpressure fraction ($\eta_O$; solid lines) for the spherical 1-D run
and the 3-D Cartesian runs with $r_{\rm cl}=0,~100$ pc; also shown in lines connected by symbols
are cumulative radiative losses. The bottom-left panel shows the time evolution of the inner and outer
shell radius for the 1-D and 3-D ($r_{\rm cl}=0$) simulations. The bottom right panel shows the angle-averaged
emissivity in the shell for the three simulations at 9.55 Myr (corresponding to the bottom panels of Fig.
\ref{fig:np-snapshots}). Note that there is a `gap' in the emissivity for the 1-D
spherical run in the dense shell where temperature is $\leq 10^4$ K and we force $\Lambda[T]=0$.
}\label{fig:1d-3d}
\end{figure*}
Three panels of Fig. \ref{fig:1d-3d} (except the bottom-right one) compare the time evolution of 1-D and 3-D simulations.
The top two panels show
that the total radiative losses are slightly higher (by $\approx 3\%$) for the 1-D spherical run
(correspondingly, mechanical energy in the box is slightly smaller), and they are similar for the
3-D Cartesian simulations with $r_{\rm cl}=0$ and $r_{\rm cl}=100$ pc. The overpressure fraction
($\eta_{\rm O}$) evolution is also very similar for the spherical 1-D and the 3-D simulation with $r_{\rm cl}=0$.
The rapid fluctuations in $\eta_{\rm O}$ at late times show that the bubble pressure is close to the ISM value and jumps above
$1.5 p_0$ after every new SN explodes inside it. The outer (inner) shell radius for the
3-D simulation (with $r_{\rm cl}=0$) is only slightly larger (smaller) than the 1-D run. To conclude,
1-D spherical simulations capture the correct evolution of  global (or volume-averaged) quantities such
as mechanical efficiency.

The bottom right panel of Fig. \ref{fig:1d-3d} shows the radial distribution of emissivity for the three runs.
For 3-D runs, average pressure and density are obtained by averaging over radial shells of size $\delta L$
and emissivity ($n_e n_i \Lambda[T]$) is calculated. The almost discontinuous rise in emissivity corresponds
to the contact discontinuity between the shocked SN ejecta and the shocked ISM. While the 1-D emissivity
profile is very sharp, the transition for 3-D runs (particularly with $r_{\rm cl}=100$ pc) is smoother. This smoothing
is due to deviation from sphericity, in particular the crinkling of the contact surface seen in the bottom panels of
Fig. \ref{fig:np-snapshots}. This also makes the shell in Cartesian simulations slightly thicker compared to the
spherical 1-D run. Radiative losses for 3-D runs are spread almost throughout the shell but are confined to
the outer radiative relaxation layer in the spherical run (see Fig. 5 in \citetalias{2014MNRAS.443.3463S}).

Both the 1-D and 3-D simulations show that the bubbles are smaller than the analytic estimates because of 
radiative cooling. Even in a uniform medium the shell can be unstable to various 3-D instabilities such as 
`Vishniac instability' \citep{1983ApJ...274..152V}, which affect the morphology of supershells (c.f. Fig.~\ref{fig:CC85-3D}; 
see also \citealt{2013A&A...550A..49K}).

\subsection{Effects of cluster \& ISM properties}\label{sec:NOB_and_n}
After discussing the fiducial run in detail, in this section we study the influence of cluster and ISM parameters
(cluster radius $r_{\rm cl}$, number of OB stars $N_{\rm OB}$, and ISM density $n_{g0}$).
\subsubsection{Effects of ISM density}
\begin{figure*}
\includegraphics[width=\linewidth]{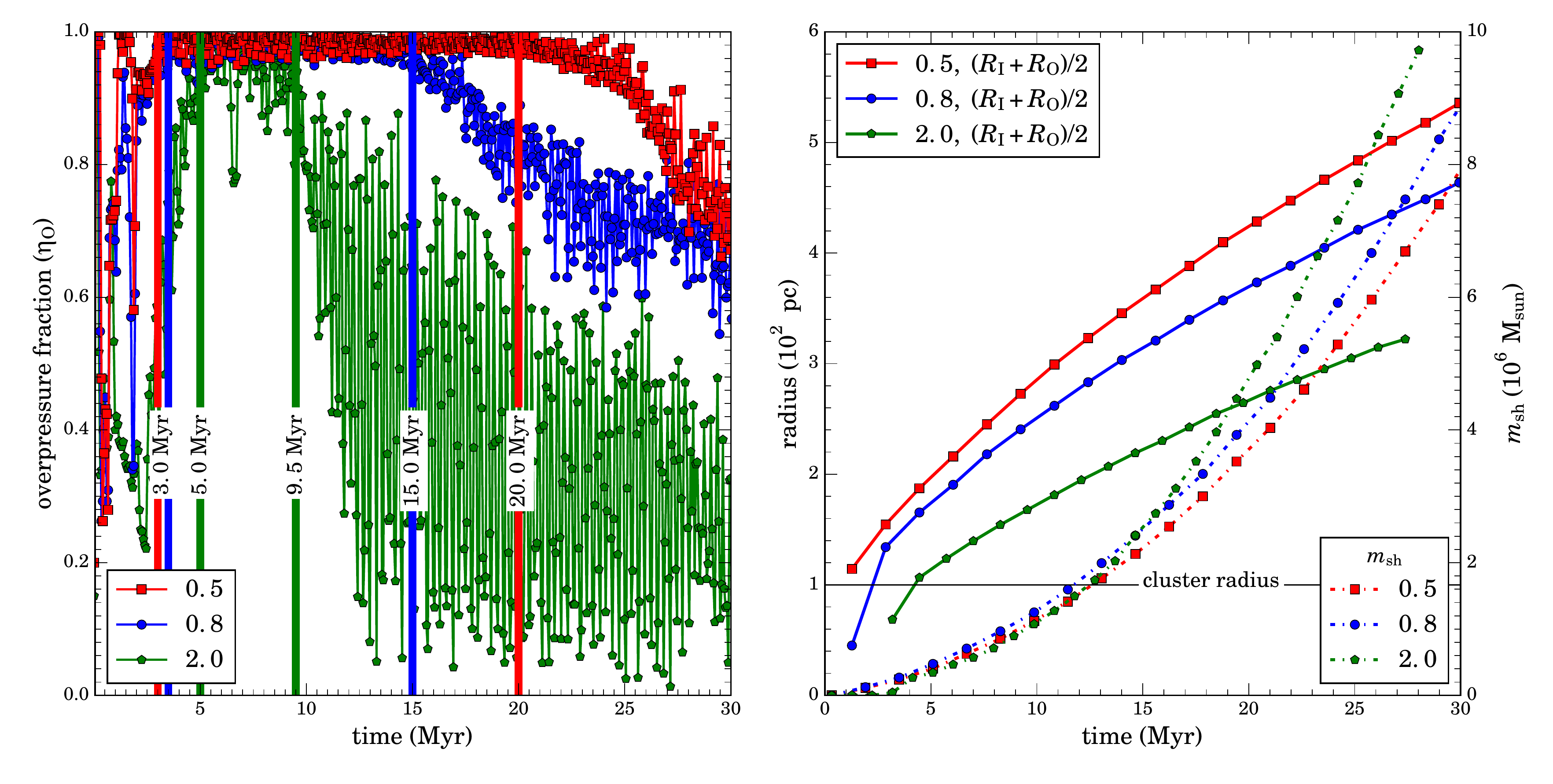}
\caption{The influence of ambient ISM density ($n_{g0}=0.5,~0.8,~2$ cm$^{-3}$) on the overpressure fraction ($\eta_O$; left panel), and the inner and
outer shell radii ($R_I,~R_O$) and the swept-up
mass in the shell ($m_{\rm sh}$; right panel) for $N_{\rm OB}=100$ and $r_{\rm cl}=100$ pc. The vertical lines in the left panel
mark the times when SNe overlap and produce
an overpressured bubble and times when they fizzle out due to radiative and adiabatic losses.
The bubble expanding in the lower density medium expands faster and sweeps up smaller mass. }
\label{fig:rad-mshell-ngas}
\end{figure*}
The gas density in which SNe explode is a crucial parameter that determines their subsequent
evolution, both in adiabatic ($r_{\rm sb} \propto \rho^{-1/5}$) and radiative (radiative losses are higher for
a larger density) regimes. The left panel of Fig. \ref{fig:rad-mshell-ngas} shows that the overpressure fraction at early times ($<5$ Myr)
both falls and rises slowly for a higher density ISM. The overlap of SNe at higher densities takes
longer because the individual bubble radius is smaller for a higher density and one needs to wait longer to
fill the whole cluster with hot gas. At late times, the overpressure fraction drops earlier for higher densities because
of larger radiative losses (although the bubble pressure scales as $n_{g0}^{3/5}$ according to \citealt{1977ApJ...218..377W}
adiabatic scaling).

The right panel of Fig. \ref{fig:rad-mshell-ngas} shows that the bubble expands more rapidly in the lower density medium.
It also shows that although the shell in a higher density ISM expands slowly, it sweeps up more mass.
An adiabatically expanding strong bubble in a uniform medium is expected to sweep up gas at a
rate $ \propto n_{g0}^{2/5}t^{9/5}$. Therefore, the ratio of mass swept by the shells with $n_{g0}=0.5,~0.8$ cm$^{-3}$
shown in Fig.~\ref{fig:rad-mshell-ngas} is expected to be
$(0.5/0.8)^{2/5} \approx 0.8$, whereas the actual value is $\approx 0.9$. This is because the bubble expanding in
a denser ISM is slower than the adiabatic model due to radiative losses; moreover, shells in a higher density medium
suffer larger radiative losses. The shell for the highest density run ($n_{g0}=2$ cm$^{-3}$) sweeps up an increasingly
larger mass at later times because $R_O \propto c_0 t$ at late times, when the shell moves close to the ISM sound speed.
\subsubsection{Effects of cluster radius}
\label{sec:rcl}
\begin{figure}
\includegraphics[width = \linewidth]{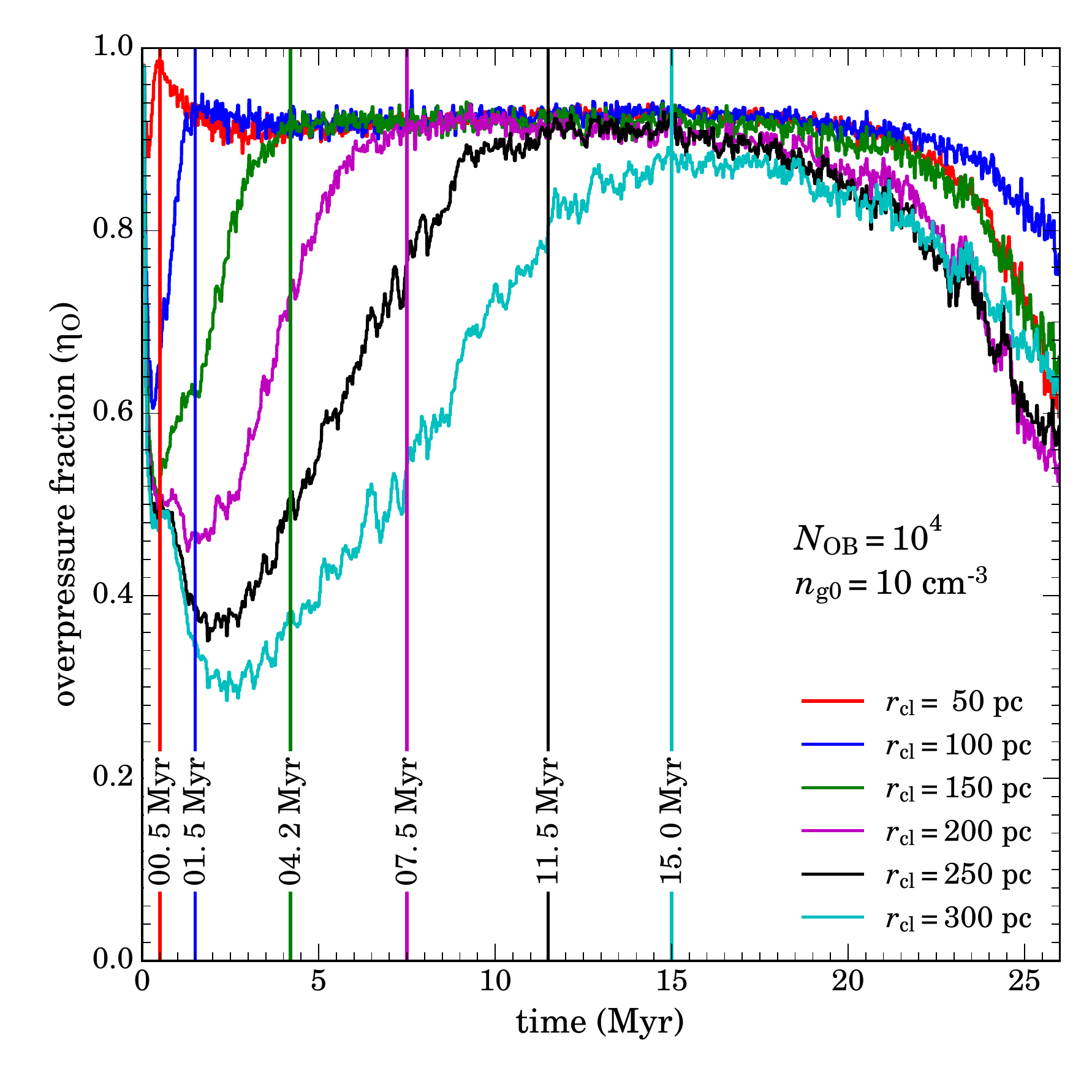}
\caption{The evolution of overpressure fraction as a function of time for $n_{g0}=10$ cm$^{-3}$
and $N_{\rm OB}=10^4$, but with different star-cluster sizes ($r_{\rm cl}$). The overpressure fraction plummets
initially as SNe are effectively isolated and cool catastrophically within 1 Myr. After that,
as more SNe go off, they start to overlap and create an overpressured bubble. As expected,
the transition to overlap happens later for a larger star-cluster. The late time drop in
overpressure fraction, occurring due to adiabatic and radiative losses, is similar for different
$r_{\rm cl}$. This suggests that the superbubble evolution is independent of the cluster size, once the coherent
overlap of SNe occurs.}\label{fig:hot-volume-fraction-vs-cluster-size}
\end{figure}

The key difference of this work from \citetalias{2014MNRAS.443.3463S} is  that we are doing 3-D simulations,
which are necessary to study a realistic spatial distribution of SNe. In 1-D spherical setup all SNe
can only explode at the origin because of spherical symmetry. Fig.~\ref{fig:hot-volume-fraction-vs-cluster-size}
shows the evolution of overpressure volume
fraction ($\eta_{\rm O}$) as a function of time for simulations with
$N_{\rm OB}=10^4$, $n_{g0}=10$ cm$^{-3}$, and different star cluster radii.\footnote{ {\scriptsize Here we choose parameters
($N_{\rm OB},~n_{g0}$) different from the fiducial run because the different stages of evolution
are nicely separated in time for this choice.The temporal behaviour is expected to be qualitatively similar for different choice of parameters.} }
The plot has a characteristic shape with an initial fall, a rise and saturation, and
an eventual fall. The initial fall occurs as isolated SNe, without
overlapping, fizzle out due to radiative losses (the top panels of Fig. \ref{fig:np-snapshots}
show density and pressure in this stage). The rise happens as SNe overlap and form a
superbubble. Eventually, the overpressure volume fraction drops as the volume of the superbubble
becomes too large and the outer shock weakens due to adiabatic and radiative losses.

We can estimate the time when SNe {\it start to overlap}. The radius of an isolated SN remnant
is given by $r_{\rm SNR} \sim (E_{\rm SN}t^2/\rho)^{1/5}$. Suppose $n_{\rm t}$ SNe have gone off independently
by some time $t$. The volume occupied by the non-overlapping SN remnants is
$\sim \sum_{i=1}^{n_t} (4\pi/3) (E_{\rm SN} i^2\delta t_{\rm SN}^2/\rho)^{3/5} \sim (4\pi/3)
(E_{\rm SN} \delta t_{\rm SN}^2/\rho)^{3/5} \sum_{i=1}^{n_t} i^{6/5} \sim (4\pi/3) (E_{\rm SN} \delta t_{\rm SN}^2/\rho)^{3/5} (5/11)
(t N_{\rm OB}/\tau_{\rm OB})^{11/5}$. Equating this volume with the volume of the star cluster $4\pi r_{\rm cl}^3/3$
gives an estimate for the time when SNe start to overlap ($t_{\rm o, ad}$, estimate for SNR overlap
assuming adiabatic evolution),
\begin{equation}
t_{\rm o, ad} \sim 0.16~{\rm Myr}~\tau_{\rm OB,30}^{5/11} N_{\rm OB,4}^{-5/11} E_{{\rm SN}, 51}^{-3/11}
n_{g0,1}^{3/11} r_{\rm cl, 2}^{15/11},
\label{eq:to_ad}
\end{equation}
where $n_{g0,1}$ is gas number density in units of 10 cm$^{-3}$ and $r_{\rm cl, 2}$ is the radius of the
star cluster in units of $100\ \rm pc$. Note that we have used Eq. \ref{eq:dtSN_tau} to obtain the above equation.

We can make another estimate for the SN overlap timescale by assuming that SNe overlap only
after they have become radiative. In this case, by a similar argument as that of the last paragraph, the
overlap time $t_{\rm o, rad}$ is given by $\tau_{\rm OB}/N_{\rm OB} (r_{\rm cl}/r_{\rm b,rad})^3$, where
$r_{\rm b, rad} \sim 37~{\rm pc}~E_{\rm SN,51}^{1/3} n_{g0}^{-1/3}$ (Eq. 2 in \citealt{2013MNRAS.434.3572R})
is the hot/dilute bubble radius when the remnant becomes radiative. Note that the bubble radius does not
increase by more than a factor of 2 after this time (e.g., Fig. 2 in
\citealt{2015ApJ...802...99K}).
Thus, the overlap time, assuming a radiative bubble, is given by
\begin{equation}
t_{\rm o, rad} \sim 0.6~{\rm Myr}~\tau_{\rm OB,30}
N_{\rm OB,4}^{-1} E_{{\rm SN}, 51}^{-1} n_{g0,1} r_{\rm cl, 2}^3.
\label{eq:to_rad}
\end{equation}
The evolution seen in Fig. \ref{fig:hot-volume-fraction-vs-cluster-size} lies somewhere in between
Eqs. \ref{eq:to_ad} \& \ref{eq:to_rad}.

The time for the overpressure volume to saturate after overlap of SNe and {\it transition to a
superbubble} evolution is given by (using \citealt{1977ApJ...218..377W} scaling and setting the superbubble
shell radius equal to the cluster radius),
\begin{equation}
t_{\rm sb} \sim 1.2~ {\rm Myr}~ r_{\rm cl, 2}^{5/3} N_{\rm OB, 4}^{-1/3} \eta_{\rm mech,-1}^{-1/3}
E_{\rm SN,51}^{-1/3} t_{\rm OB,30}^{1/3} n_{g0, 1}^{1/3},
\label{eq:t_sb}
\end{equation}
where we have scaled the result with a mechanical efficiency $\eta_{\rm mech}$ of 0.1 (i.e., only
$\sim10\%$ of the input SN energy goes into blowing the superbubble; $\sim 90\%$ is lost
radiatively). This estimate for the time of superbubble formation roughly matches the results
in Fig. \ref{fig:hot-volume-fraction-vs-cluster-size}. Finally, the time when the superbubble pressure ($\sim 0.75 \rho \varv_{\rm sb}^2$)
falls to $\approx1.5$ times the ISM pressure is given by (apart from factors of order unity, this is
essentially the same as Eq. \ref{eq:tiff})
\begin{equation}
t_{\rm fiz} \sim 10.3~{\rm Myr}~T_4^{-5/4} \eta_{\rm mech,-1}^{1/2} E_{\rm SN, 51}^{1/2}
\tau_{\rm OB}^{-1/2} N_{\rm OB,4}^{1/2} n_{g0,1}^{-1/2},
\label{eq:t_fizzle}
\end{equation}
($T_4$ is the ISM temperature in units of $10^4$ K) which is only slightly lower than the time corresponding to
the late time drop in the overpressure volume fraction in Fig.~\ref{fig:hot-volume-fraction-vs-cluster-size}. Note that unless the cluster
size ($r_{\rm cl}$) is unrealistically large, overlap of supernovae is likely to occur. In this state the time for a superbubble to fizzle out
is independent of the cluster size.
\subsubsection{Effects of supernova rate: formation of a steady wind}
\citealt{1985Natur.317...44C} found a solution (hereafter CC85) for the wind driven by internal energy
and mass deposited uniformly within an injection radius ($r<R$). This was applied to the galactic outflow
in M82. For a large number of SNe (i.e., a large $N_{\rm OB}$), the mechanical energy injection can
be approximated as a constant luminosity wind,
$L_w=N_{\rm OB}E_{\rm SN}/\tau_{\rm OB}$. According to CC85, within the injection radius ($r \lesssim R$)
the mass density is constant,
whereas at large radii (wind region, $r \gtrsim R$) density is expected to be $\propto r^{-2}$.
A termination shock is expected at the radius where the wind ram pressure balances
the pressure inside the shocked ISM. For small $N_{\rm OB}$, however, the individual SN ejecta does not
thermalize within the termination shock radius ($r_{\rm TS}$) as the SN occurs inside a
low density bubble (the bubble density is low in the absence of significant mass loading as
most of the ambient gas is swept up in the outer shell) created by the previous SNe.
For a large SN rate the solution should approach the steady state described
by \citealt{1985Natur.317...44C}. \citetalias{2014MNRAS.443.3463S} derived analytic constraints
on $N_{\rm OB}$ required for the existence of a smooth CC85 wind inside the superbubble (see their Eq. 11)
as,
\begin{equation}
\delta t_{\rm SN,CC85 }\gtrsim 0.008~{\rm Myr}~E_{\rm SN,51}^{-9/26}t_{\rm Myr}^{4/13}
n_{g0}^{-3/13}M_{\rm SN,5\odot}^{15/26},
\label{eq:CC85}
\end{equation}
where $M_{\rm SN,5\odot}$ is the SN ejecta mass and $t_{\rm Myr}$ is the age of the starburst in Myr.
This time between SNe corresponds to a requirement of $N_{\rm OB} \gtrsim 4 \times 10^3$ for a
smooth CC85 wind to appear by 1 Myr. Using the standard stellar mass function, this corresponds to a
star formation rate of $\sim 0.01 M_\odot {\rm yr}^{-1}$. This is a lower limit because thermalization just before
the termination shock does not lead to a high density/emissivity core, the characteristic feature of a CC85 wind.
\begin{figure}
\includegraphics[width=\linewidth]{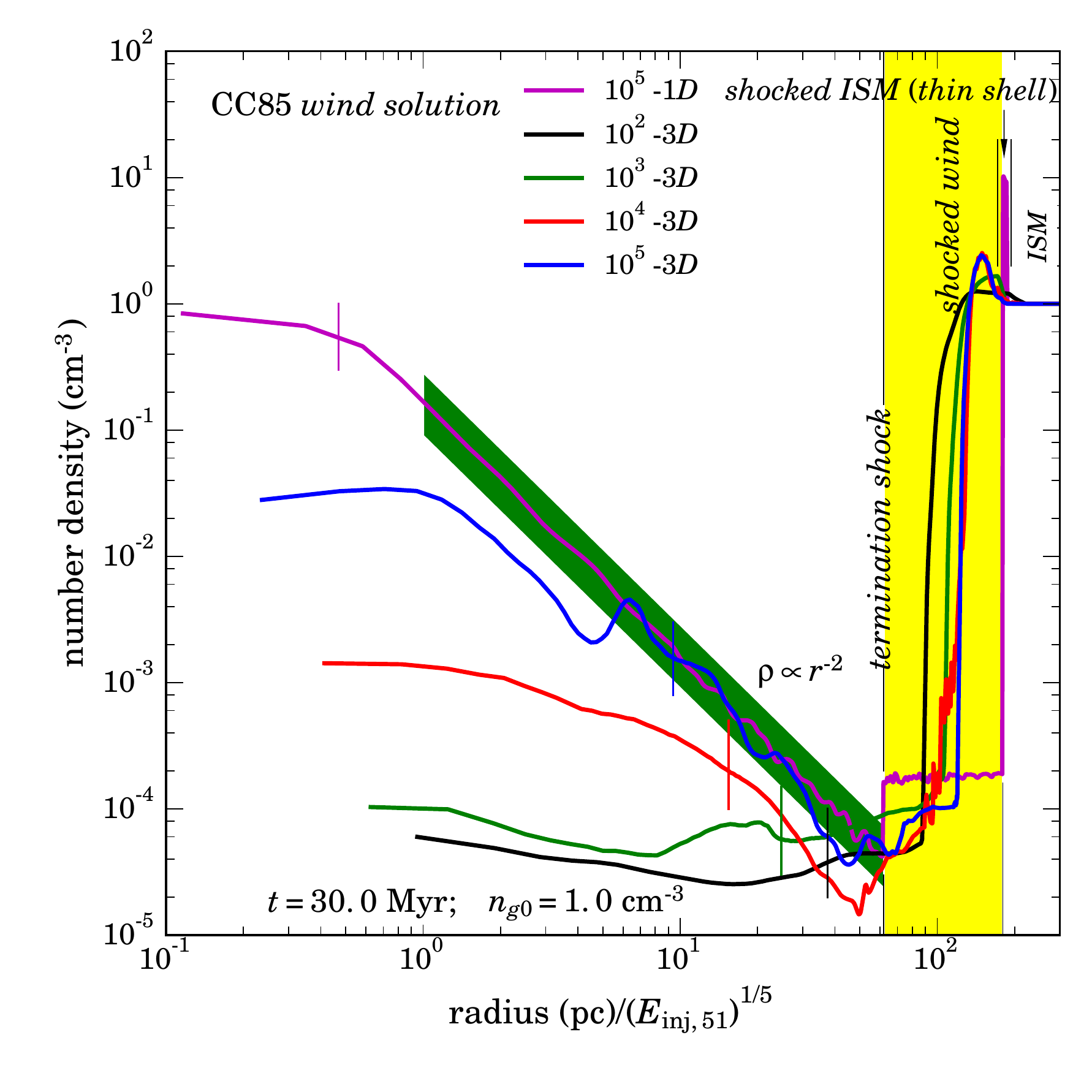}
\caption{Spherically averaged gas density profiles for 3-D runs at 30 Myr ($r_{\rm cl}=100$ pc, except for the 1-D spherical run;
$n_{g0}=1$ cm$^{-3}$)
with various $N_{\rm OB}$. Radius has been scaled with the expected scaling
( $E_{\rm inj,51}^{1/5} \approx N_{\rm OB}^{1/5}$ is the total mechanical energy injected in units of $10^{51}$ erg; see Table
\ref{table:simulations}). The thin vertical lines mark the cluster radius ($r_{\rm cl}$) in the scaled unit. The density profile attains
a smooth, steady CC85 profile (its signature is the $\rho \propto r^{-2}$ profile
beyond a core region) within the bubble for a large $N_{\rm OB} \gtrsim 10^4$,
consistent with the analytic considerations in section 4.3 of \citetalias{2014MNRAS.443.3463S} (see also Fig. 3 in their paper).
The radiative shell in 1-D run (with the same resolution as the 3-D run) is much thinner as compared to 3-D because the 3-D shell
is not perfectly spherical and the contact discontinuity is crinkled (see Fig. \ref{fig:CC85-3D}). The outer shock is weaker for a
smaller $N_{\rm OB}$ but its location scales with the analytic scaling ($\propto N_{\rm OB}^{1/5}$).}
\label{fig:CC85}
\end{figure}
Fig~\ref{fig:CC85} shows the density profiles for a range of $N_{\rm OB}$
($N_{\rm OB}=10^5$ corresponds to a SN rate of $\sim 0.003 \ \rm yr^{-1}$). As expected from
thermalization of a SN within the ejecta of all previous SNe (Eq. \ref{eq:CC85}), a smooth CC85-type wind
with density $\propto r^{-2}$ at 30 Myr only forms for $N_{\rm OB} \gtrsim 10^4$. Since SNe form in OB associations,
they are expected to overlap and form superbubbles. For a sufficiently large number of SNe ($\gtrsim 10^5$; e.g., in the
super star clusters powering a galactic wind in M82) a strong termination shock (with Mach number $\gg 1$) exists till late times,
which may accelerate majority of Galactic and extragalactic high energy cosmic rays (e.g., \citealt{2004A&A...424..747P}).
In contrast, strong shocks (especially the reverse shock; \citealt{1974ApJ...188..335M})
in isolated SNe exist only at early times ($\lesssim 10^3$ yr), after which the reverse shock crushes the central neutron star
and the outer shock weakens with time (in fact catastrophically after it becomes radiative).
\begin{figure}
\includegraphics[width=1.1\linewidth]{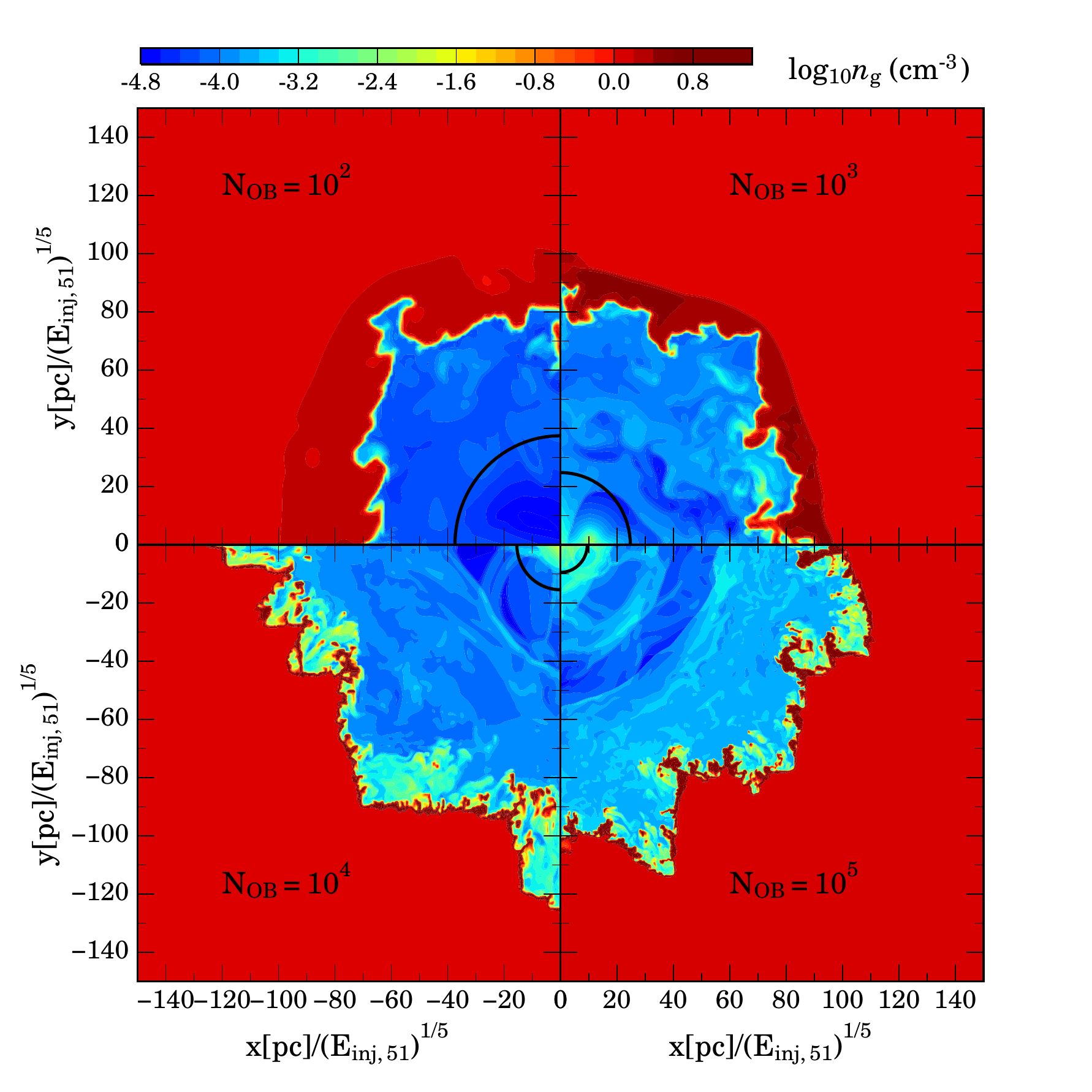}
\caption{Density slices at $t=9.55$ Myr in the $x-y$ ($z=0$) plane for runs with different $N_{\rm OB}$
but with the same ISM density and cluster radius ($n_{g0}=1$ cm$^{-3}$ and $r_{\rm cl}=100$ pc).
The shell becomes progressively thinner with increasing $N_{\rm OB}$
because a stronger shock causes higher compression. Like Fig. \ref{fig:CC85}, the axes are scaled with respect
to the analytic scaling. The solid black arcs mark the cluster radius. A termination shock and a high density injection region
are visible in the bottom two panels.}
\label{fig:CC85-3D}
\end{figure}
Fig. \ref{fig:CC85-3D} shows the 2-D density snapshots of the 3-D runs shown in Fig. \ref{fig:CC85}, albeit at an earlier time.
As expected, the shell is much thinner for a larger number of SNe. Also, a dense injection region and a clear termination shock
are visible for the runs with $N_{\rm OB} \gtrsim 10^4$. Crinkling of the contact discontinuity and the thin shell is the key
difference of 3-D runs as compared to the spherical 1-D simulations.

The SNe driven wind is able to maintain a strong non-radiative termination shock that is able to power the outward motion of
the outer shock. The CC85 model has two parameters: the efficiency with
which star formation is converted into thermal energy ($\alpha \equiv \dot{E}/{\rm SFR}$), and the mass loading factor
($\beta \equiv \dot{M}/{\rm SFR}$), which determine the properties of galactic outflows (e.g., \citealt{2016ApJ...818L..24S}).
From our setup we can determine the mass-loading for large $N_{\rm OB}$ simulations by calculating the mass loss rate
from the cluster measured at radii where the mass outflow rate $\dot{M}(r) \equiv 4\pi r^2\rho \varv$ is roughly constant.
The mass loading factor for our $N_{\rm OB}=10^5$ run is $\approx 1$ as most of the SN injected mass flows out in a
roughly steady wind. For much larger $N_{\rm OB}$ (or equivalently, SFR) valid for starbursts, the mass loading factor can
be reduced because of radiative cooling and mass drop-out from the dense ejecta of SNe (e.g., 
\citealt{2007IAUS..237..497W,2008ApJ...683..683W,2011ApJ...740...75W}). \citealt{2016MNRAS.456.3432G} have investigated
launching of galactic outflows based on multi-physics simulations which include variation in SN rate and various strategies for 
placing supernovae (random, or at density peaks or isolated). \citealt{2016ApJ...824L..30P,2016ApJ...827L..29S} investigate the
effect of cosmic ray diffusion on dynamics of galactic outflows. We will investigate the effect of additional processes in
our future work.
\section{Discussion}\label{sec:discussion}
In this section we discuss the astrophysical implications of our work, focusing on radiative losses,
comparison the observed HI supershells, and gas expulsion from star clusters.
\subsection{Mechanical efficiency \& critical supernova rate for forming a superbubble}\label{sec:eta_nob_ng}
\begin{figure}
\includegraphics[width=1.1\linewidth]{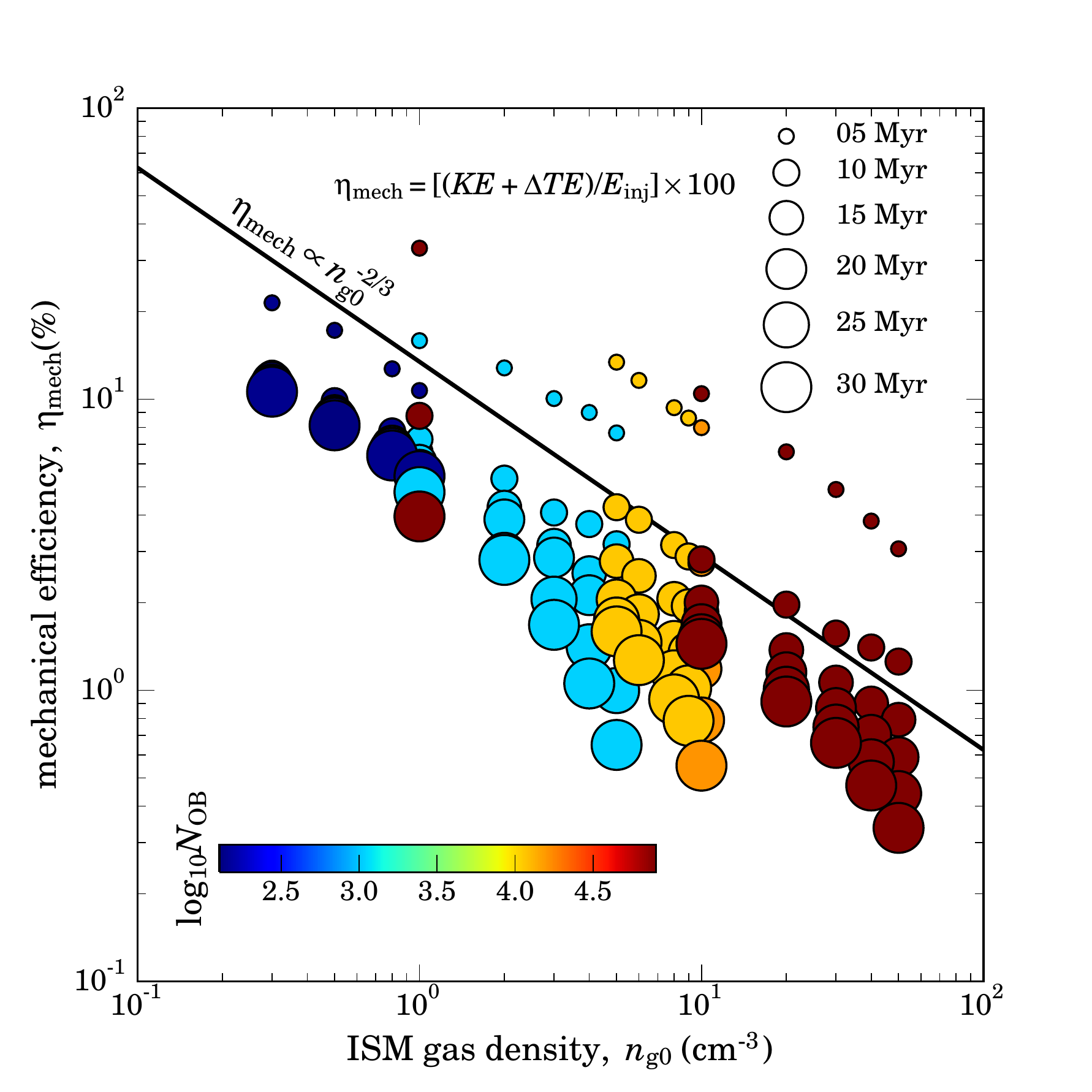}
\caption{Efficiency of mechanical energy retention in superbubbles as a function of ISM density
for different simulations. Clearly, the efficiency decreases with an
increasing density because of radiative losses. Colors represent $N_{\rm OB}$ and the sizes of
circles stand for different times. Note that for the same density,
a higher $N_{\rm OB}$ gives a slightly larger mechanical efficiency. The black solid line shows the
$n_{g0}^{-2/3}$ scaling of $\eta_{\rm mech}$, which describes well the variation of mechanical
efficiency with the ambient density at almost all times. For most runs, especially with high ambient densities ($n_{g0} \gtrsim 2$ cm$^{-3}$),
the mechanical efficiency decreases by almost a factor of 10 from 5 to 30 Myr.}
\label{fig:efficiency}
\end{figure}

While isolated SNe lose {\it all} their energy by $\sim 1\ \rm Myr$, even overlapping
SNe forming superbubbles lose majority of energy injected by SNe.
The mechanical efficiency of superbubbles is defined as
\begin{equation}
\eta_{\rm mech} \equiv \frac{(KE + \Delta TE )}{E_{\rm inj} },
\label{eq:mech_eff}
\end{equation}
where $KE$ is the total kinetic energy of the box, $\Delta TE$ is the increase in the box
thermal energy, and $E_{\rm inj}$ is the energy injected by SNe (which is slightly
larger than $N_{\rm OB}E_{\rm SN}$ because mass is added at the local velocity).
By energy conservation (the computational box is large enough that energy is not
transported in to or out of it), $\eta_{\rm mech} = 1 - RL/E_{\rm inj}$, where $RL$ are
cumulative radiative losses. Fig. \ref{fig:efficiency} shows the mechanical efficiency
(Eq. \ref{eq:mech_eff}) as a function of the initial gas density ($n_{g0}$) at various times for runs
with different $N_{\rm OB}$. One immediately sees that mechanical efficiency decreases
with an increasing ISM density ($n_{g0}$). Efficiency also decreases with time (by almost
a factor of $10$ from 5 to 30 Myr), especially for higher densities. The maximum efficiency
is $\sim 20\%$, occurring at early times. Our simulations show that the mechanical efficiency
of 3-D and 1-D simulations are comparable and almost independent of the cluster size
($r_{\rm cl}$, see section \ref{sec:1D} \& Fig. \ref{fig:1d-3d}), provided that SNe overlap before fizzling out.
A rough scaling of $\eta_{\rm mech} \propto n_{g0}^{-2/3}$, valid at most times, can be deduced from Fig. \ref{fig:efficiency}.
Also note that the mechanical efficiency increases very slightly for a larger number of SNe.

Fig. \ref{fig:efficiency} shows mechanical efficiencies that are about an order of magnitude smaller
than the values quoted in \citetalias{2014MNRAS.443.3463S}. For example, the efficiency (which
equals $1-$ fractional radiative losses; see the right panel of Fig. 8 in \citetalias{2014MNRAS.443.3463S})
for $N_{\rm OB}=10^5$ and $n_{g0}=1$ cm$^{-3}$ in \citetalias{2014MNRAS.443.3463S} at 30 Myr is
$\approx 40\%$. The value for the same choice of parameters from Fig. \ref{fig:efficiency} is $\approx 6\%$,
smaller by a factor of $\approx 7$.  This discrepancy is mainly due to the much higher resolution in the 1-D 
simulations of \citetalias{2014MNRAS.443.3463S} (see section \ref{sec:eta_conv}). 
\begin{figure}
\includegraphics[width=1.1\linewidth]{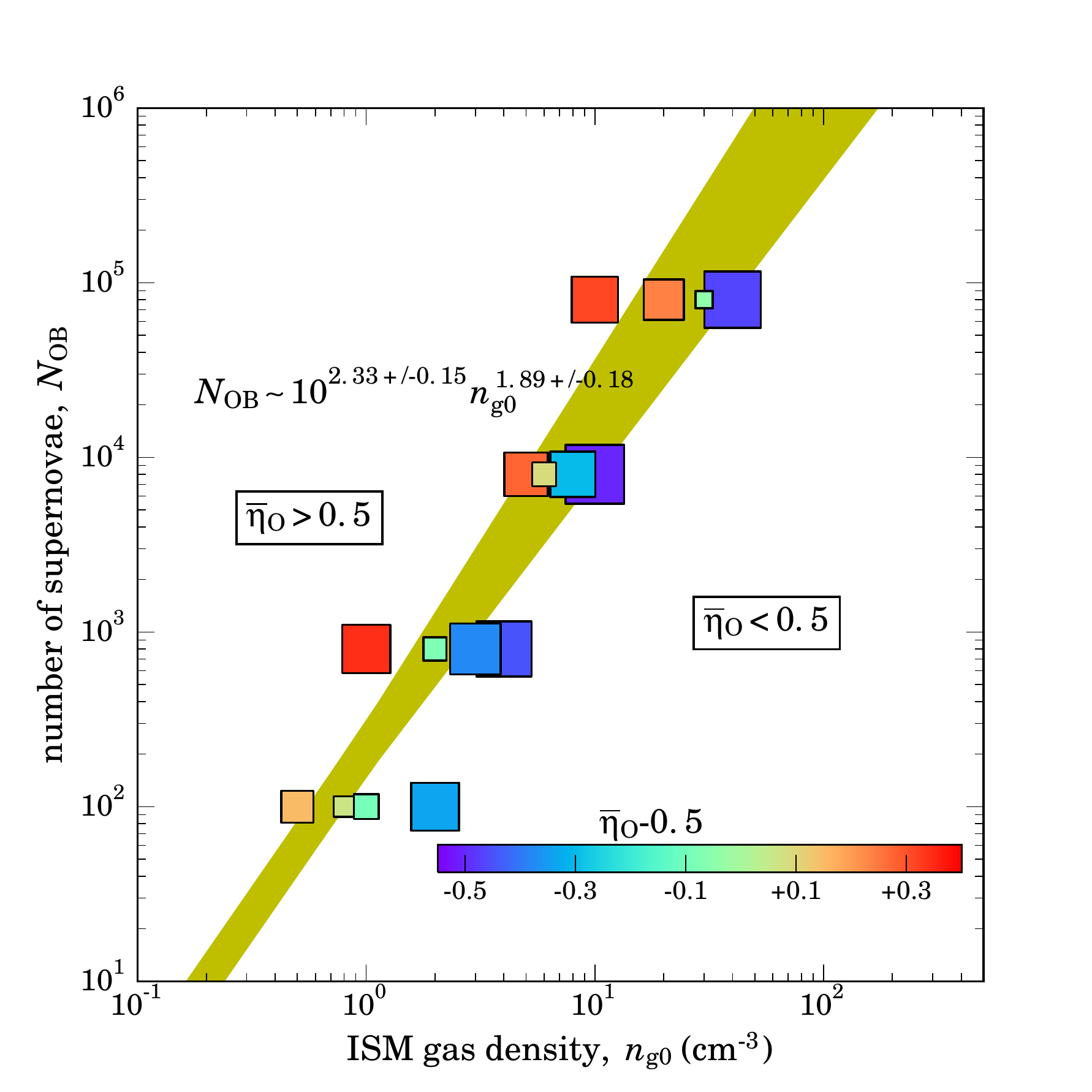}
\caption{Critical $N_{\rm OB}$ required for a given density of the ISM for the superbubble
to remain sufficiently overpressured at late times. The colorbar and the size of squares (a smaller
square means that $\bar{\eta}_O$ is closer to 0.5)
represent deviation from an average overpressure fraction of 0.5 at late times (25 to 30 Myr).
The best-fit power-law scaling is indicated, and $3-\sigma$ spread about the best fit is indicated
by the shaded region.}
\label{fig:critical}
\end{figure}

Fig. \ref{fig:hot-volume-fraction-vs-cluster-size} shows that the overpressure volume
fraction $\eta_O$ for $n_{g0}=10$ cm$^{-3}$ and $N_{\rm OB}=10^4$ has a similar value
for cluster sizes as large as $r_{\rm cl} = 300$ pc. This means that the evolution of the superbubble is independent
of $r_{\rm cl}$, as long as overlap of SNe happens
before the cluster age, which is very likely not only for individual star clusters but also
for clusters of star clusters as in the center of M82 galaxy
(\citealt{1995ApJ...446L...1O}). Therefore, the key parameter that determines if the
superbubble remains sufficiently overpressured by the end of the star-cluster
lifetime, for a given gas density, is the number of SNe $N_{\rm OB}$
(and not the cluster size $r_{\rm cl}$).

The overpressure volume fraction (defined in Eq. \ref{eq.volfrac}) is an appropriate diagnostic
to determine if a superbubble has fizzled out or not. As described earlier, we consider a superbubble
fizzled out if the average overpressure fraction falls below 0.5 at late times (25 to 30 Myr).
Fig. \ref{fig:critical} shows the plot of critical number of SNe required to produce
an average overpressure volume fraction of 0.5 at late times (25 to 30 Myr), for a given gas
density. We vary the ISM density for a given $N_{\rm OB}$, such that the late-time overpressure
fraction is close to 0.5. The critical $N_{\rm OB}$ roughly scales as $n_{g0}^2$.

Now we turn to analytic arguments to understand the scaling of critical $N_{\rm OB}$ for a given
ISM density ($n_{g0}$). The superbubble pressure as a function of time, according
to the adiabatic model of \citet{1977ApJ...218..377W}, is $\sim \frac{3}{4} \rho \varv_{\rm sb}^2$,
which at the end of cluster lifetime becomes
\begin{equation}
\frac{p_{\rm sb, late}}{k_B} \sim 1.7 \times 10^5~{\rm K~cm}^{-3} N_{\rm OB,4}^{2/5}
\eta_{\rm mech,-2}^{2/5} \tau_{\rm OB, 30}^{-6/5} n_{g0,1}^{3/5},
\label{eq:pressure_late}
\end{equation}
where mechanical efficiency has been scaled to 0.01. Equating this to 1.5 times the ambient ISM
 pressure $p_{\rm ISM}/k_B = 10^5 n_{g0,1} T_4$, gives
\begin{equation}
N_{\rm OB,crit} \sim 7.3\times 10^3 \eta_{\rm mech,-2}^{-1} \tau_{\rm OB,30}^3 n_{g0,1} T_4^{5/2}.
\label{eq:NOBcrit}
\end{equation}
This estimate of the critical number of OB stars to maintain an overpressured bubble at late times agrees with
Fig.~\ref{fig:critical} in that the critical $N_{\rm OB}$ for
$n_{g0}=10$ cm$^{-3}$ is about $10^4$.  From Eq. \ref{eq:NOBcrit}, we get the scaling of critical $N_{\rm OB}$ as
$N_{\rm OB, crit} \propto n_{g0} \eta_{\rm mech}^{-1} $, which when we use the dependence of
$\eta_{\rm mech}$ on $n_{g0}$ from Fig. \ref{fig:efficiency}
($\eta_{\rm mech} \propto n_{g0}^{-2/3}$), gives $N_{\rm OB, crit} \propto n_{g0}^{5/3}$. This scaling
is similar to the scaling of critical $N_{\rm OB}$ observed in Fig.
\ref{fig:critical}; namely,
\begin{equation}
\label{eq:crit_curve}
N_{\rm OB, crit} \approx 200 n_{g0}^{1.89}  \tau_{\rm OB,30}^3 T_4^{5/2}.
\end{equation}
A steeper $\eta_{\rm mech}$ versus $n_{g0}$, which is not inconsistent
with Fig. \ref{fig:efficiency}, will give an even better match. The important
point to note is that a decreasing mechanical efficiency with an increasing ISM
density, is required to explain the critical $N_{\rm OB}$ curve.

The scaling between $N_{\rm OB}$ and gas density (hereafter critical curve)
in Fig.~\ref{fig:critical} can be compared with
the empirical relation between star formation rate (SFR) and gas density. The Kennicutt-Schmidt (hereafter KS) relation
\citep{1959ApJ...129..243S,1998ApJ...498..541K} between gas surface density and SFR surface density is
\begin{equation}
\label{eq:KS}
{ \Sigma_{\rm {SFR}} \over {\rm M}_\odot \, {\rm yr}^{-1} \, {\rm kpc}^{-2}} \approx
3 \times 10^{-3} \left ( { \Sigma_g  \over 10 \, {\rm M}_\odot \, {\rm pc} ^{-2}} \right )^{1.4} \,,
\end{equation}
which is valid for $\Sigma_g \ge 10$ M$_\odot$ pc$^{-2}$,
below which a much steeper relation holds \citep{2008AJ....136.2846B}.
Consider a scale height ($H$) of $100$ pc and a disk radius ($R_d$) of 1 kpc. For each OB star, the
total stellar mass is $\sim 100$ M$_\odot$ for
Kroupa/Chabrier initial mass function \citep{2002Sci...295...82K,2003PASP..115..763C}.
Then for a star formation time scale of 30 Myr, we have
$\Sigma_{\rm SFR} \approx 10^{-6} N_{\rm OB}$ M$_\odot$ yr$^{-1}$ kpc$^{-2}$ $\tau_{\rm OB,30}^{-1} R_{d, {\rm kpc}}^{-2}$.
For a gas density of $n_{g0}$ cm$^{-3}$, we also have
$ \Sigma_g \approx 3 \, n_{g0} H_{\rm 100 pc}$ M$_\odot$ pc$^{-2}$ for mean molecular weight $\mu=1.3$
(assuming neutral/molecular disk).
Therefore, the KS relation can be re-written in terms of the parameters used in this paper as
\begin{equation}
N_{\rm OB, KS} \approx 550 \, n_{g0}^{1.4} H_{\rm 100 pc}^{1.4} R^2_{d,{\rm kpc}} \tau_{\rm OB,30}.
\label{eq:NOB_KS}
\end{equation}

An important point to note is that the scaling of $N_{\rm OB, KS}$ with gas density is somewhat {\it shallower}
than the scaling for the critical curve (Eq. \ref{eq:crit_curve}), but comparable in magnitude (this depends on the assumed scale height
and disk radius).
In the case of starbursts, the normalization for the KS relation can be larger
\citep{2012ARA&A..50..531K},
but this normalization is also consistent with the critical curve in this paper. Also note that the critical number of OB stars (Eq.
\ref{eq:crit_curve}) depends sensitively on the ISM temperature and can be much
smaller for a cooler (say 100 K) disk. Therefore, a comparison of Eqs. \ref{eq:crit_curve} \& \ref{eq:NOB_KS}
should be made only after using appropriate disk/ISM parameters. A steeper slope for critical $N_{\rm OB}$
as compared to the KS relation (despite the dependence on other parameters as disk radius and ISM temperature)
implies that SNe can disrupt the star-forming regions more easily in weak/moderate star-forming regions
but not in dense starbursts. This may explain the observed higher efficiency of star-formation (or a larger
normalization of KS relation; Eq. \ref{eq:KS}) in starbursts relative to moderate star forming regions. The key
uncertain step in this argument (which is beyond the scope of this paper) is how the maintenance of an overpressured
bubble translates into suppression of star formation.

We can also compare our critical $N_{\rm OB}-n_{g0}$ curve with the observed threshold of SFR $\sim 0.1$ M$_\odot$ yr$^{-1}$ kpc$^{-2}$
for galactic superwinds (e.g., \citealt{2002ASPC..254..292H}). Using similar arguments used to derive Eq. \ref{eq:NOB_KS},
the critical SFR density corresponding to our critical curve is
\begin{equation}
\Sigma_{\rm {SFR, crit}} \sim 2 \times 10^{-4} {\rm M}_\odot {\rm yr}^{-1} {\rm kpc}^{-2} n_{g0}^{1.89} R_{d, \rm kpc}^{-2}.
\label{eq:Heckman}
\end{equation}
While this is much smaller than the Heckman limit, it is comparable to the lower limit on SFR density for the appearance of
radio halos in spiral disks, $\sim 10^{-4}$ erg cm$^{-2}$s$^{-1}$ (equivalent to SFR density of
$10^{-5} {\rm M}_\odot {\rm yr}^{-1} {\rm kpc}^{-2}$; \citealt{1995ApJ...444..119D}). \citet{2013MNRAS.434.3572R} argue
that to form a galactic superwind the superbubble has to break out with a sufficiently high Mach number ($\gtrsim 5$), but
our critical curve (Eq. \ref{eq:NOBcrit}) is based on a Mach number of unity.
\subsection{Radius - velocity distribution of HI supershells}
\begin{figure}
\includegraphics[width=1.05\linewidth]{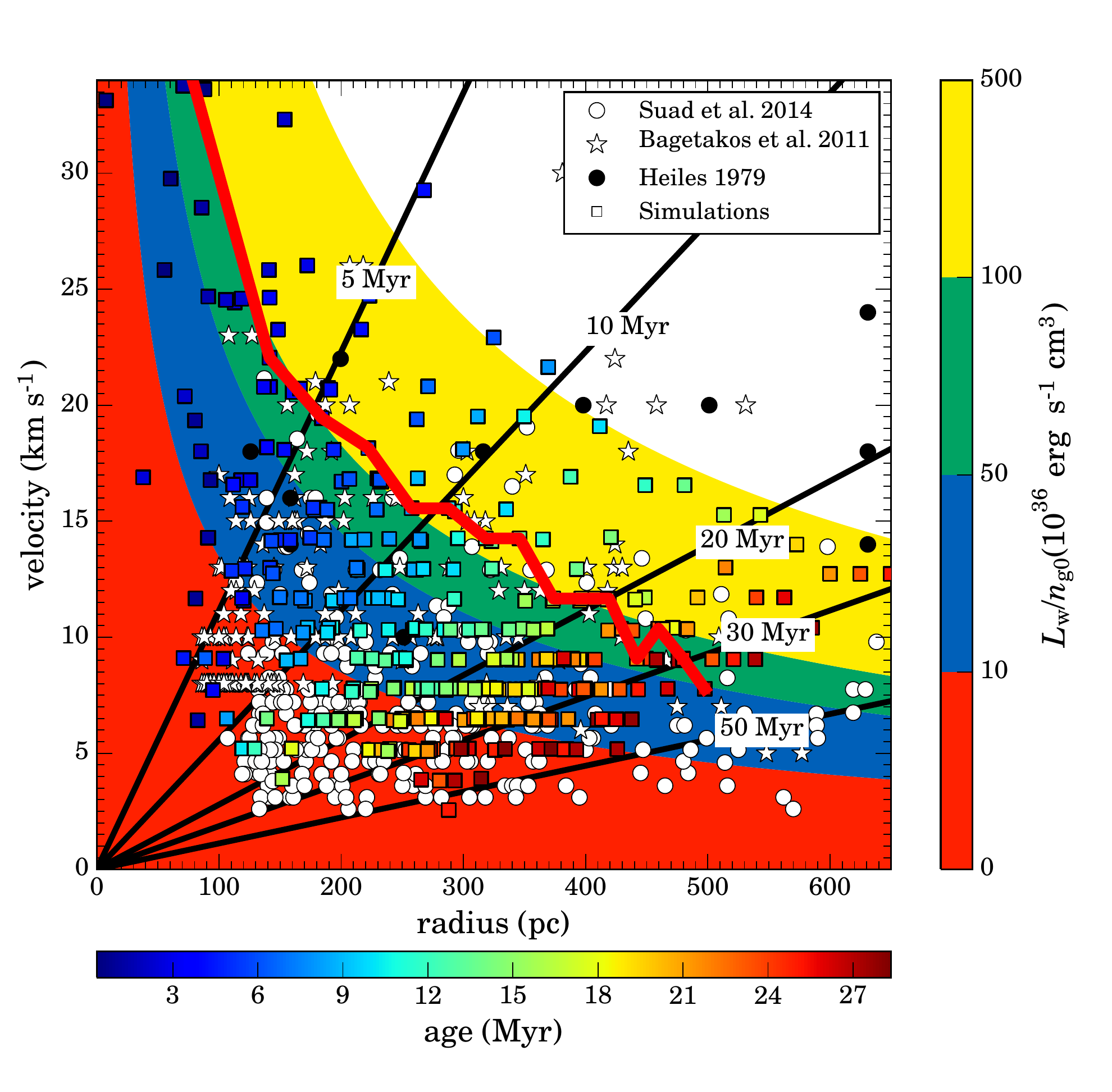}
\caption{Radius-velocity distribution of supershells from analytic estimates (vertical colorbar; $\eta_{\rm mech}=1$ is used in Eq. \ref{eq:v_sb}),
our numerical simulations (colored squares; horizontal colorbar indicates time), and observations
(black and white symbols; \citealt{1979ApJ...229..533H,2014A&A...564A.116S,2011AJ....141...23B}) of HI supershells. The radius from
simulations corresponds to the inner radius in spherically averaged density profiles ($R_I$) of the shell and the velocity is obtained by $dR_I/dt$.
Solid black lines correspond to the dynamical age ($1.67r_{\rm sb}/\varv_{\rm sb}$; see Eqs.  \ref{eq:r_sb} \&  \ref{eq:v_sb}). The solid red line
shows the evolution of velocity and radius for $N_{\rm OB}=1000$, $n_{g0}=1$ cm$^{-3}$, $r_{\rm cl}=100$ pc. At early times the line closely
follows the analytic curve for $L_w/n_{g0}=10^{38}~{\rm erg~s^{-1}}$ (corresponding to $N_{\rm OB}=100$ and $n_{g0}=1$ cm$^{-3}$), with a factor
 $\sim 10$ smaller mechanical luminosity; at later times it dips even further. This is consistent with the radiative efficiency of $\lesssim 10\%$,
which decreases with time (Fig. \ref{fig:efficiency}).}
\label{fig:r-v-plot}
\end{figure}
Our setup provides an opportunity to study the observed properties of HI shells supershells,
especially the ones which are close to spherical and not much affected by the background density
stratification. Fig~\ref{fig:r-v-plot} compares the radius-velocity distribution of observed HI supershells
\citep{1979ApJ...229..533H,2014A&A...564A.116S,2011AJ....141...23B} with the evolution seen in our numerical
simulations; also shown are the \citet{1975ApJ...200L.107C} analytic scalings.
We can write the radius and velocity of the shell in terms of the parameter $L_{\rm w, 38}/n_{g0}$
(luminosity $L_{\rm w, 38} \approx N_{\rm OB,2}E_{SN,51}/\tau_{\rm OB,30} $ is the mechanical luminosity scaled to $10^{38}$ erg s$^{-1}$)
using Eq.~\ref{eq:v_r_sb} as
\begin{equation}
\frac{L_{\rm w, 38}}{n_{g0}} =
\left(\frac{r_{\rm sb}}{58\ \rm pc}\right)^{2}\left(\frac{\varv_{\rm sb}}{34\ \rm km\ s^{-1}}\right)^{3}.
\end{equation}
The vertical colorbar in Fig. \ref{fig:r-v-plot} shows the contours of constant $L/n_{g0}$. The `$\square$' symbols
represent the radius and velocity obtained from our simulations ($N_{\rm OB}/n_{g0}$ ranges from 100 to $10^5$); it is
very encouraging that the observed distribution of $r_{\rm sb}-\varv_{\rm sb}$ is similar to our simulations, which correspond to reasonable
star cluster parameters.
The solid red color track marks the evolution of a bubble with
$N_{\rm OB} \approx 10^3$ and $n_{g0}=1.0\ \rm cm^{-3}$, which corresponds to
$L_{\rm w, 38}/n_{g0} \approx 10$. But the track lies close to the analytic contours of
$L_{\rm w, 38}/n_{g0} \approx 0.5-1.0$. It means that, for a given $r_{\rm sb}-v_{\rm sb}$ the adiabatic theory
overestimates $L_{\rm w, 38}/n_{g0}$ by a factor of $\sim 10-20$. This discrepancy is primarily due to large radiative
losses; mechanical efficiency in Fig. \ref{fig:efficiency} $\lesssim 10\%$ is consistent with the evolution in the
$r_{\rm sb}-\varv_{\rm sb}$ space. Also we note that some of the simulation points (below the $t=30\ \rm Myr$ line)
have a dynamical age ($\equiv 5r_{\rm sb}/3v_{\rm sb}$) longer than the simulation time.

If SNe are the dominant cause of bubble formation, then we require large OB associations for the
creation of the observed HI supershells. In order to quantify the size of OB associations, we
also need to evaluate the mechanical energy injection from stellar winds and radiation. However,
even without accounting for these additional energy/momentum sources, the observed shells are much
smaller and slower compared to what is expected from the predictions of adiabatic theory applied to
the observed stellar population -- the so-called power problem in superbubbles (\citealt{2009AIPC.1156..295O} and
references therein). Our simulations show that radiative losses can account for the power problem.

\subsection{Gas removal from clusters}
\begin{figure*}
\includegraphics[width=\linewidth]{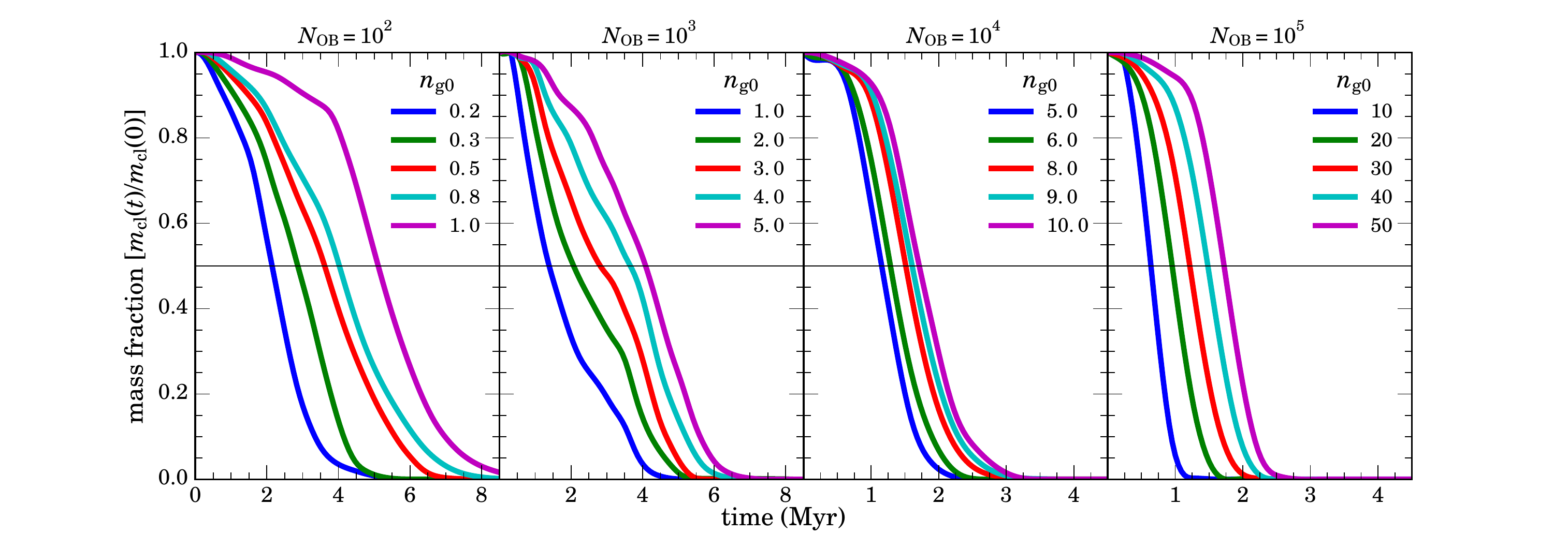}
\caption{The fraction of original cluster gas remaining within the cluster ($r<r_{\rm cl}$; $r_{\rm cl}=100$ pc for all these runs)
as a function of time for various SN counts ($N_{\rm OB}$). Each of the sub-panels show curves corresponding to different values
of gas density $n_{g0}$, the initial ISM density inside the cluster. It is noted that even a small $N_{\rm OB}$ makes the cluster
lose all its gas by about $ 8-10\ \rm Myr$. The evacuation time increases with density, as expected.}
\label{fig:gas_expulsion}
\end{figure*}
Due to the presence of feedback from OB stars (radiative, stellar winds and SNe) the star
forming regions clear gas on timescales $\sim 10^{6}\ \rm yr$ \citep{2003ARA&A..41...57L}. For
clusters simulated by us the $m_{\rm stars}/m_{\rm cl}\sim 0.17 N_{\rm OB,2}/n_{g0}r_{\rm cl,100}^3<1$ (for each OB star,
the total stellar mass is $\sim 100$ M$_\odot$ for Kroupa/Chabrier initial mass function), therefore the gravitational well
is largely provided by the cluster gas (this is also true for embedded clusters buried in their natal molecular clumps). As a result of gas expulsion the cluster potential
becomes shallower and the cluster may become unbound depending on the ratio of gas removal
timescale and dynamical timescale of the cluster \citep{2003ARA&A..41...57L}. If the gas expulsion
time is long compared to the dynamical time, the stars can adiabatically attain new viral equilibrium
without being unbound. However, in the opposite regime because of a suddenly reduced gravity
majority of stars become unbound. The timescale of gas expulsion is also important to account for
multiple populations observed in globular clusters (e.g., \citealt{2016A&A...587A..53K} and references therein).

While our simulations do not account for gravity that holds the star cluster together (inclusion of gravity is
important for strongly bound massive clouds, not so much for smaller clumps with lower gravitational
binding energies), we can qualitatively understand the action of supernova/stellar wind energy injection in
gas expulsion from star clusters. Fig.~\ref{fig:gas_expulsion}  shows the mass fraction
$m_{\rm cl}(t)/m_{\rm cl}(0)$ ($m_{\rm cl}$ is the gas mass inside the cluster radius, $r<r_{\rm cl}$) as a function of time for
various values of $N_{\rm OB}$ and $n_{g0}$. Since the ratio of energy injected by SNe
to the gravitational potential energy $\sim N_{\rm OB}E_{\rm SN}/(G \mu^2 m_p^2 n_{g0}^2 r_{\rm cl}^5)
\sim 5\times 10^5 N_{\rm OB,2} E_{\rm SN, 51}/(n_{g0}^2 r_{\rm cl, 2}^5)$
is large, the effect of neglecting gravity is negligible for the choice of our parameters (for simulations with gravity, see
\citealt{2015ApJ...814L..14C,2016A&A...587A..53K}).
We find that the clusters are evacuated due to the formation of a superbubble within
$\lesssim 10\ \rm Myr$ (Fig.~\ref{fig:gas_expulsion}).  As expected,  lower ISM density and higher $N_{\rm OB}$
evacuate the cluster gas in a shorter time. An estimate of evacuation timescale is given in Eq. \ref{eq:t_sb}. The estimate
depends strongly on the cluster radius ($r_{\rm cl}$) but is weakly sensitive to parameters such as ISM density, $N_{\rm OB}$,
$\eta_{\rm mech}$, etc. The results in Fig. \ref{fig:gas_expulsion} are consistent with the timescale in Eq. \ref{eq:t_sb}; therefore,
for different parameters our numerical results can be scaled according to the theoretical scaling.
\subsection{Convergence of $\eta_{\rm mech}$ \& temperature distribution of radiative losses}\label{sec:eta_conv}
\begin{figure}
\includegraphics[width = \linewidth]{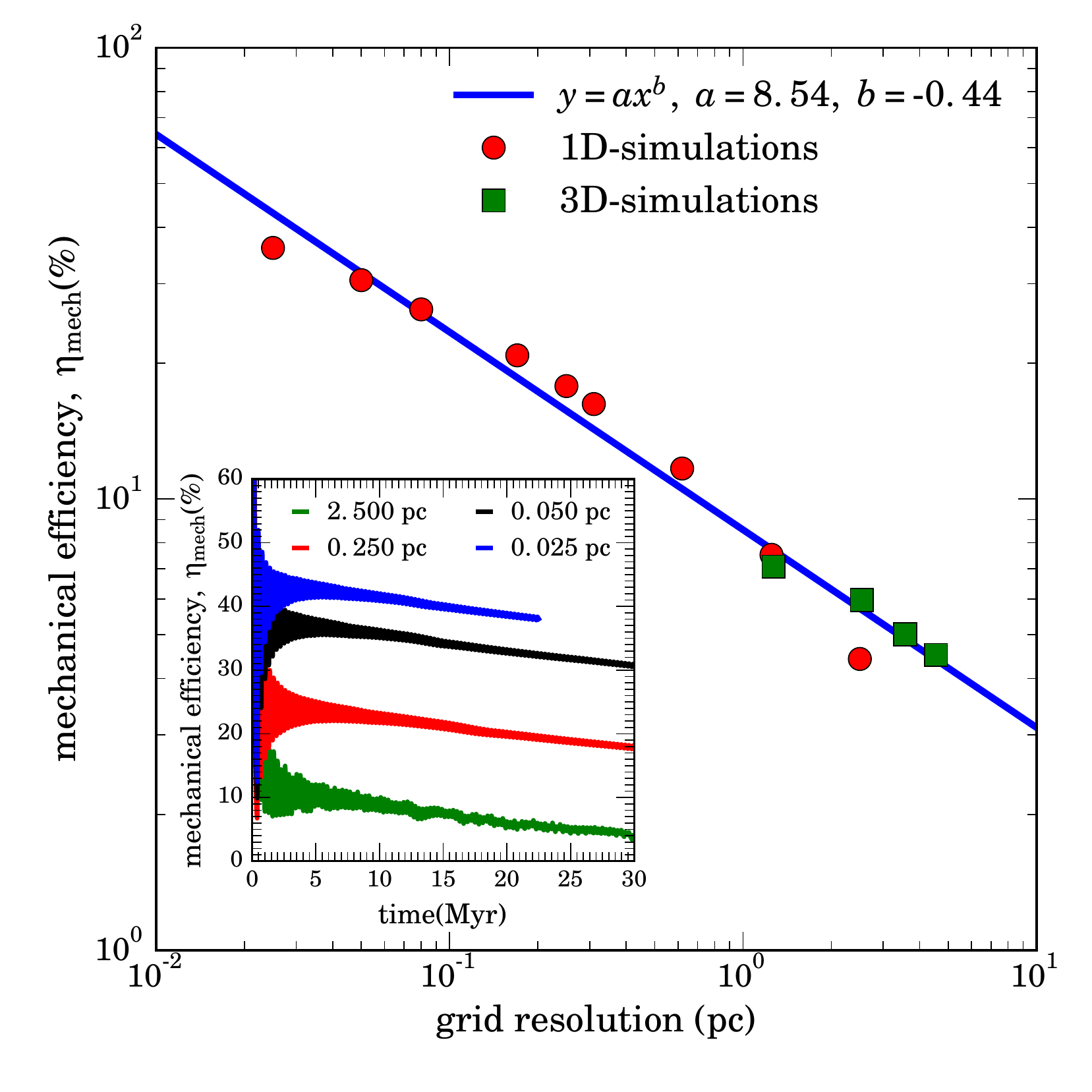}
\caption{Mechanical efficiency measured at 30 Myr as a function of grid resolution for various 1-D and 3-D superbubble 
simulations. The grid parameters are $N_{\rm OB}=100$, $n_{\rm g0}=1\ \rm cm^{-3}$ and $r_{\rm cl}=0$ (for 1-D runs) and 
$r_{\rm cl}=100$ pc (for 3-D runs). The blue solid line is the best least squares power-law fit to the data points. The inset shows
the evolution of mechanical efficiency for the high resolution 1-D runs. Mechanical efficiency does not converge even for the 
highest  resolution 1-D simulations.
}\label{fig:efficiency-grid-res}
\end{figure}

One of the key questions is whether our results are converged. Convergence of the fiducial 3-D simulation is discussed in
Appendix \ref{sec:convergence}. Fig. \ref{fig:convergence} clearly shows that the higher resolution simulations show finer
features. What about the convergence of volume-averaged quantities such as mechanical efficiency ($\eta_{\rm mech}$)?
Fig. \ref{fig:efficiency-grid-res} shows mechanical efficiency (Eq. \ref{eq:mech_eff}) measured at 30 Myr
for the fiducial 3-D and 1-D runs at various resolutions. Even average quantities like $\eta_{\rm mech}$ do not show perfect 
convergence (we get a higher value of $\eta_{\rm mech}$ with increasing resolution). The 1-D simulations can be carried out at a 
much higher resolution than the 3-D ones, and yet $\eta_{\rm mech}$ increases with an increasing resolution. In section 
\ref{sec:1D} we show that at the same resolution the radiative losses are comparable in 3-D and 1-D (top-right panel of 
Fig. \ref{fig:1d-3d}). From this, we expect that even the very high resolution 3-D simulations (which are beyond the capabilities 
of current computational resources) will not show convergence.

Recent very high resolution 1-D simulations (\citealt{2016arXiv160601242G,2016MNRAS.462.4532G}) have highlighted the importance
of very high resolution to obtain mechanical efficiency and momentum delivered to the ISM by supernovae. However,
our Fig. \ref{fig:efficiency-grid-res} clearly shows the lack of convergence even at the highest resolutions. 
The cooling losses in any simulation with unresolved boundary layers (radiative relaxation layer and contact discontinuity) 
will keep on decreasing with an increasing resolution because the volume of cooling layers (and hence radiative loss rate) decreases 
with an increasing resolution. This means that convergence can only be achieved by explicitly including diffusive
processes such as thermal conduction and/or viscosity, which can numerically resolve the radiative layers. Moreover, the values of
physical conductivity and viscosity are too small (especially for the dense phases) to be resolved on the grid. Therefore, artificially large
numerical diffusivities (which may crudely mimic small-scale turbulent transport) must be used. The importance of
resolving cooling layers via explicit thermal conduction to obtain convergence in thermal instability simulations is highlighted in \citet{2004ApJ...602L..25K}. 
Similarly \citet{2007A&A...476.1113F,2007MNRAS.378.1471L} show that explicit resistivity and viscosity are required to get converged
results for angular momentum transport due to magnetorotational instability (MRI) in unstratified shearing boxes.

\begin{figure}
\includegraphics[width=\linewidth]{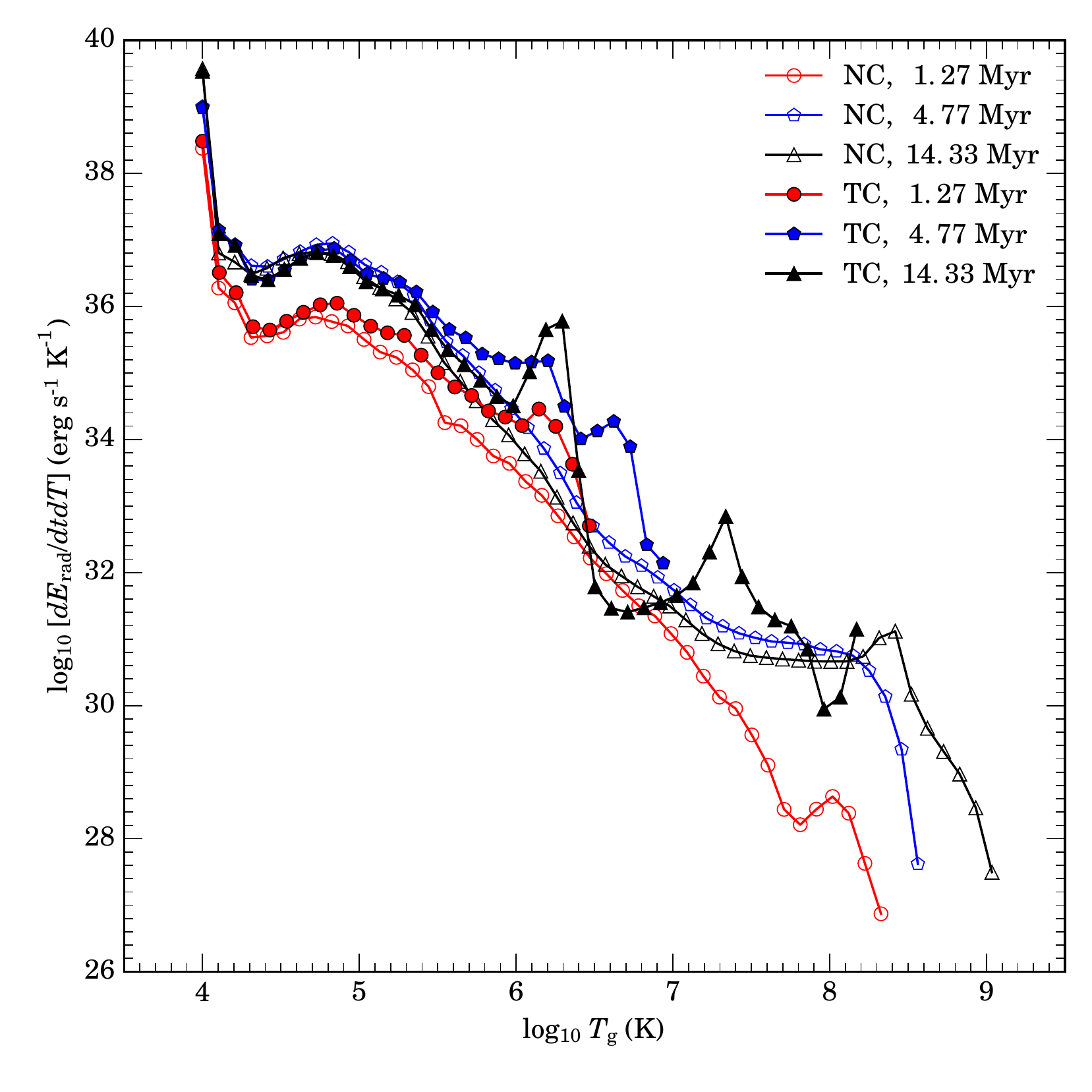}
\caption{Radiative loss rate per temperature bin ($dE_{\rm rad}/dtdT$) as a function of gas temperature at different 
times for our fiducial simulation with (TC) and without (NC) thermal conduction. 
We calculate the radiative loss rates in logarithmically
spaced temperature bins with $\Delta  \log_{10} T=0.1$. Thermal conduction reduces the maximum gas temperature in the box 
because of evaporation of matter into the hot bubble.}
\label{fig:temp_luminosity}
\end{figure}

One observationally important diagnostic is the temperature distribution of cooling losses in superbubbles; this determines
the wavebands in which they emit. Fig.~\ref{fig:temp_luminosity} shows the temperature distribution of the radiative loss rate for the fiducial 3-D
run with and without thermal conduction. Both with and without conduction, the radiative losses occur primarily at $\sim 10^4$ K; the fractional radiative
losses for $T<10^5$ K are 99.6\% and 99.3\% with and without conduction, respectively. 
This result is consistent with the recent superbubble simulations in dense molecular gas \citep{2016MNRAS.462.4532G}, which show that the cooling 
losses at $\sim10^4$ K are about two order of magnitude larger than X-ray ($\sim 10^{6-7}$ K) and molecular ($\sim 100$ K) losses.
Thermal conduction reduces the maximum temperature in the hot bubble due to the evaporation of mass
from the dense shell to the hot bubble, as shown in Fig.~\ref{fig:conduction-flow}.
\section{Conclusions} \label{sec:conclusions}
We have carried out 3-D hydrodynamic simulations of supernovae (SNe) in an OB association
that creates and drives a superbubble. Our aim has been to study the effect of
multiple SNe distributed over a limited region of a cluster, on the ambient material far
outside the cluster, and derive the dependence of fundamental parameters such as
the efficiency of energy deposition and the critical number of SN required to create overpressured
bubbles. Our settings have been admittedly, and intentionally, kept idealized
so that we can perform controlled numerical experiments. Physical effects such as magnetic fields,
thermal conduction, stratification, and inhomogeneities in the ambient gas, which we have not included here,
presumably do play important roles in superbubble formation and evolution, and will be the focus of our future
studies.

The broad astrophysical implications of our results are discussed in section \ref{sec:discussion}.
Our key results can be summarized as follows:
\begin{itemize}
\item While isolated SNe fizzle out by $\sim 1$ Myr due to radiation losses,
for a realistic cluster
size it is likely that subsequent SNe go off in a hot and tenuous medium and
sustain a shock lasting for the cluster lifetime $\sim 30$ Myr, comparable to the galactic
dynamical timescale. 1-D numerical simulations faithfully capture the global energetics
but cannot, by construction, capture morphological features such as the crinkling of the
contact discontinuity seen in 3-D.
\item  While most of the input energy is lost via radiative cooling, the superbubble
retains a fraction $\eta_{\rm mech}$ of the input energy, and this fraction scales as $\eta_{\rm mech} \propto n_{g0}^{-2/3}$,
being of order $\sim 6\%$ for $n_{g0}\sim 1$ cm$^{-3}$ over a time period of $\sim 30$ Myr. We note that
the mechanical efficiency increases with an increasing resolution, and that converged result can only be obtained
by resolving cooling layers using explicit diffusion.
\item We have explored the parameter space of ISM density ($n_{g0}$), number of
SNe ($N_{\rm OB}$) and star cluster radius ($r_{\rm cl}$) to study the conditions for the
formation of an overpressured superbubble. For realistic cluster sizes, we find that
the bubble remains overpressured only if, for a given $n_{g0}$, $N_{\rm OB}$ is
larger than a threshold value. Our results show that threshold condition can be roughly expressed as
$N_{\rm OB,crit} \sim 200  n_{g0}^{1.9}$, where $n_{g0}$ is the particle density in cm$^{-3}$.
\item Classical adiabatic superbubble evolution overestimates the ratio of the wind luminosity and
the ISM density ($L_{\rm w}/n_{g0}$) by a factor of $\sim 10-20$, by not taking
radiation losses into account. This explains the `power problem' of the observed size and speed
of superbubbles, and our simulations confirm that radiative losses are the reason for discrepancies
between the size-speed distribution of HI supershells and the sizes of OB associations driving them.
\item We confirm that a minimum value of $N_{\rm OB} (\gtrsim 10^4)$ is need to produce a steady wind
and a strong termination shock within the cluster region. For a smaller number of SNe, all the supernova
energy is deposited at the radiative dense shell.
\end{itemize}
\section*{Acknowledgments}
We are grateful to the Supercomputing Education and Research Centre (SERC) at IISc for facilitating
our use of Cray XC40-SahasraT cluster, without which these challenging simulations could not be carried out.
This work is partly supported by the DST-India grant no. Sr/S2/HEP-048/2012 and an India-Israel
joint research grant (6-10/2014[IC]; which also supports Naveen Yadav). We wish to thank Kartick Sarkar, Arpita Roy,
Deovrat Prasad, Prakriti Pal Choudhury and Prasun Dhang for many helpful discussions. We thank Siddhartha
Gupta for convincing us to study the convergence of mechanical efficiency in 1-D.
\bibliographystyle{mnras}
\bibliography{references}

\appendix
\section{Radius determination of the shell}
\label{app:rad}
\begin{figure}
\includegraphics[width=\linewidth]{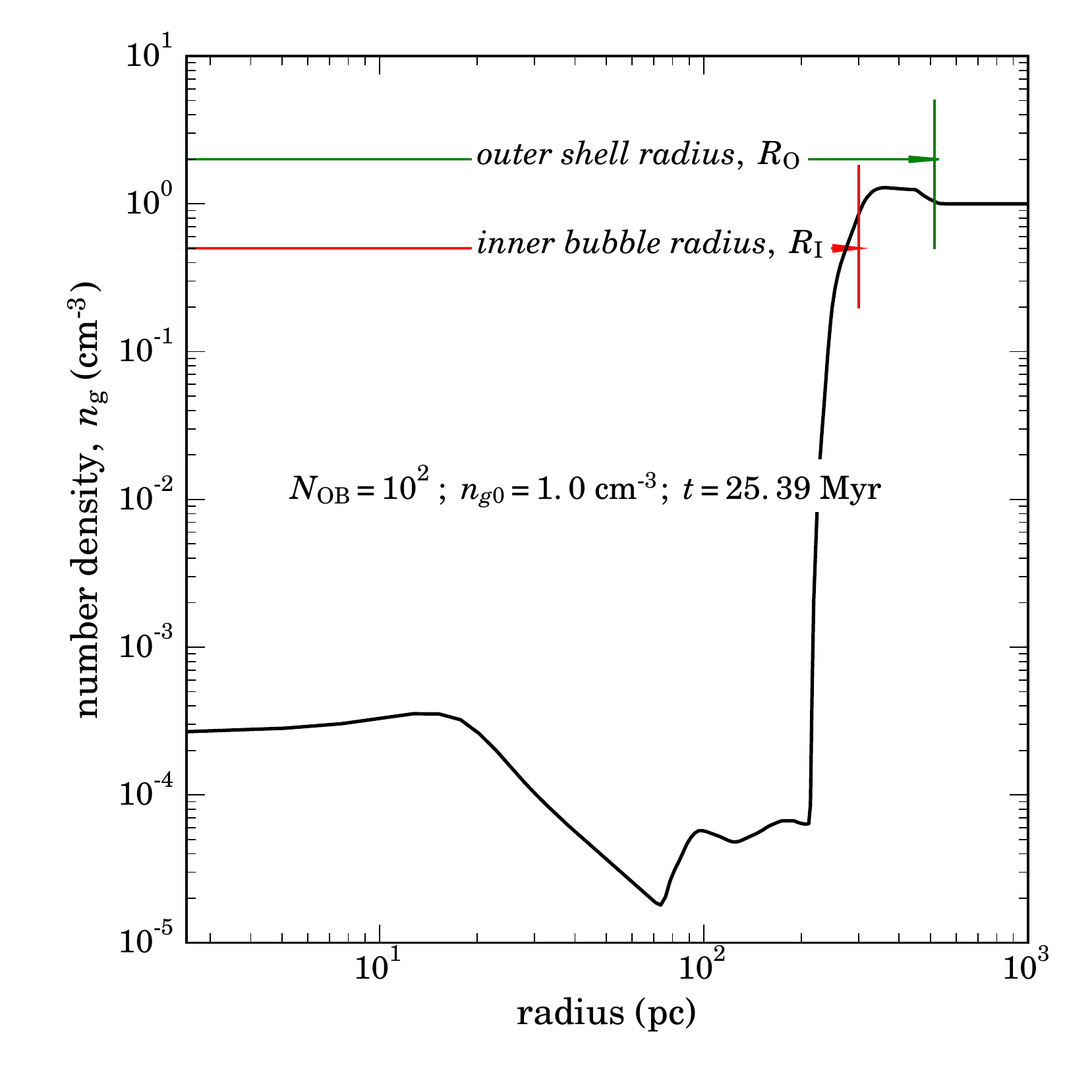}
\caption{Determining the radius of supershells: first we calculate the angle-averaged density profiles
in spherical shells of size $\delta r=\delta L$. The outer shell radius ($R_O$) corresponds to the radius at which
the average density is larger than $1.02$ times the ambient ISM density ($n_{g0}$) and the inner radius ($R_I$)
corresponds to the radius at which the average density falls below $0.98n_{\rm g0}$.}
\label{fig:bubble-radius-demo}
\end{figure}

For 3-D simulations the dense shell is not perfectly spherical. Some figures (e.g., Figs. \ref{fig:rvst_n_opf}, \ref{fig:r-v-plot})
show the evolution of the shell radius with time.
Fig. \ref{fig:bubble-radius-demo} shows how we determine
the inner and outer radii of the supershells. We construct angle-averaged radial
density profiles by dividing the simulation box into spherical shells of thickness $\delta r=\delta L$, and averaging
over all the grid cells contained within the shell. The inner shell radius is taken at the radius where
$n_{\rm g}=0.98n_{\rm g0}$ and the outer shell radius has $n_{\rm g}=1.02n_{\rm g0}$.
\section{Convergence}\label{sec:convergence}
In order to ensure the convergence of the results, we carried out our fiducial run ($n_{g0}=1$ cm$^{-3}$, $T_0=10^4$ K,
$N_{\rm OB}=100$, $r_{\rm cl}=100$ pc) with different grid
resolutions (see Table~\ref{table:convergence}). The timestep is shorter for a
higher grid resolution as $\Delta t \propto N^{-1}$, where $N$ is the number of grid points along any
direction. Hence, the total computational cost scales $ \propto N^4$, which becomes prohibitive
for a large number of grid-points. An optimum resolution, large enough to capture key physical features
but computationally feasible, needs to be chosen.

\begin{figure*}
    \centering
        \includegraphics[width=0.9\textwidth]{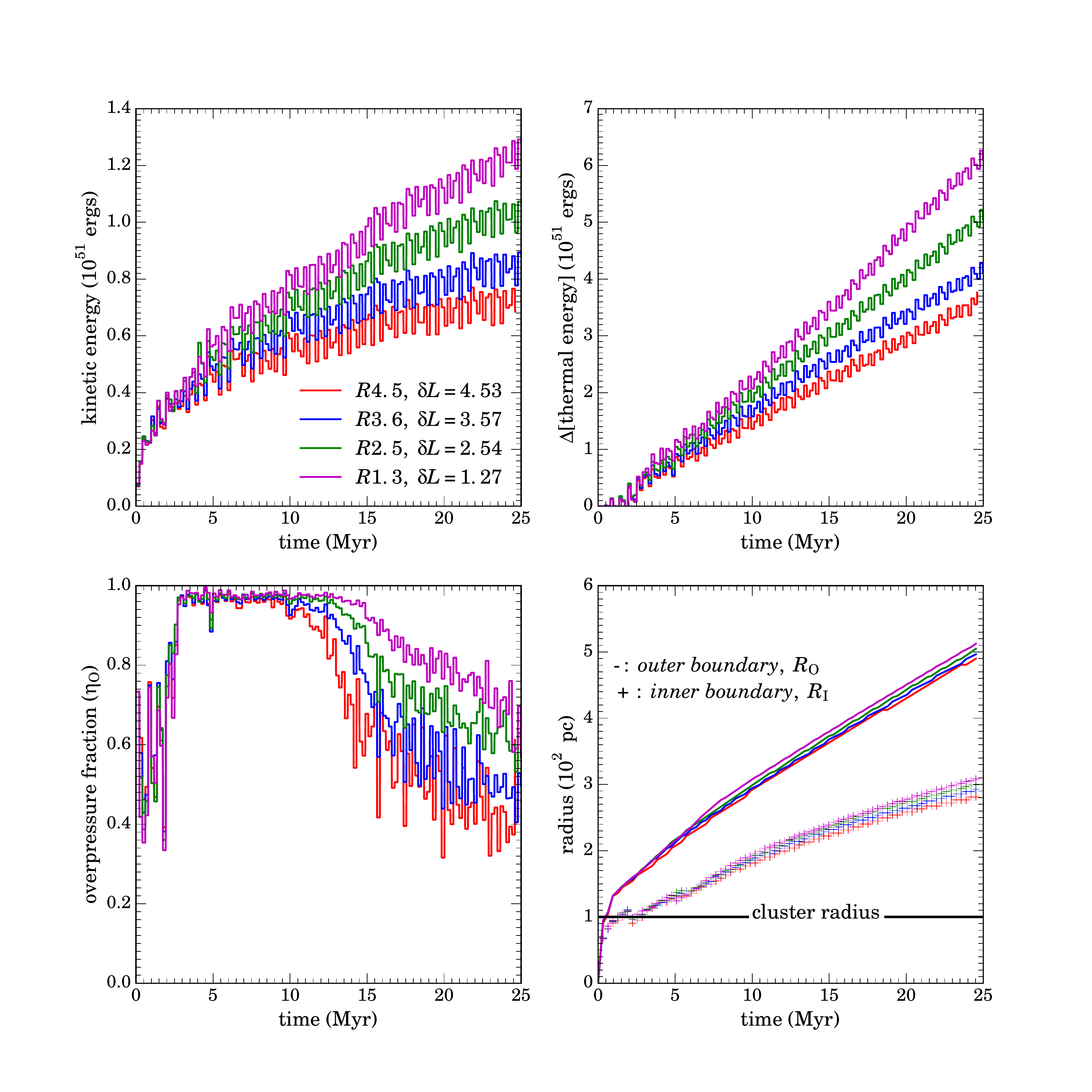}
    \caption{Various volume-averaged quantities (kinetic energy, change in thermal energy, overpressure fraction, and inner and outer radii of the shell)
    for the fiducial parameters
    ($N_{\rm OB}=100$, $n_g=1$ cm$^{-3}$, $r_{\rm cl}=100$ pc) as a function of time for different grid resolutions,
    $\delta L=1.27,~2.54,~3.57,~4.53$ pc. The top two and the bottom-left panels show binned data with a bin-size of 0.18 Myr.
    The results show convergence with an increasing resolution. }\label{fig:convergence}
\end{figure*}

Fig. \ref{fig:convergence} compares the evolution of volume integrated quantities and the shell radius
for various grid resolutions. A larger energy is retained and the overpressure fraction (Eq. \ref{eq.volfrac}) is larger
for a higher resolution, but the difference is small for the highest resolutions ($\delta L=1.27,~2.54$ pc). The evolution of the
inner and outer shell radii are also similar.
\begin{table*}
\caption{Convergence runs for fiducial parameters}
\begin{centering}
 \begin{tabular}{r r r r c c c c c c c}
 \hline
 Label  & $L$  &$N$ &$\delta L$ & $R_O^\dag$ & $R_I^\dag$ & $KE$                 & $\Delta TE$          & $E_{\rm inj}^\ddag$        & $\eta_{\rm mech}$ & $\eta_{\rm O}$  \\
        & (pc) &    & (pc)      & (pc)        & (pc)        & $(10^{51}\ \rm erg)$ & $(10^{51}\ \rm erg)$ & $(10^{51}\ \rm erg)$ & ($\%$)              &                 \\
 \hline
 R4.5  &714 & 315 &  4.54 &281 &490 &  0.69 &  3.84 & 100.38 &  4.51 &  0.40 \\
 R3.6  &714 & 400 &  3.57 &293 &496 &  0.79 &  4.37 & 103.05 &  5.01 &  0.59 \\
 R2.5  &649 & 512 &  2.54 &299 &505 &  0.98 &  5.29 & 105.02 &  5.97 &  0.68 \\
 R1.3  &649 & 1024 &  1.27 &308 &512 &  1.19 &  6.32 & 106.11 &  7.08 &  0.72 \\
 \hline
 \end{tabular}
 \end{centering}
 \begin{flushleft}
 {\scriptsize $^\dag R_{\rm O}$ ($R_{\rm I}$) is the outer (inner) radius of the shell at $\approx 25\ \rm Myr$.\\
 $^\ddag$ Kinetic and thermal energy added to the simulation box by SNe.
 }
 \end{flushleft}
\label{table:convergence}
\end{table*}

\begin{figure*}
\includegraphics[width = 0.85\linewidth]{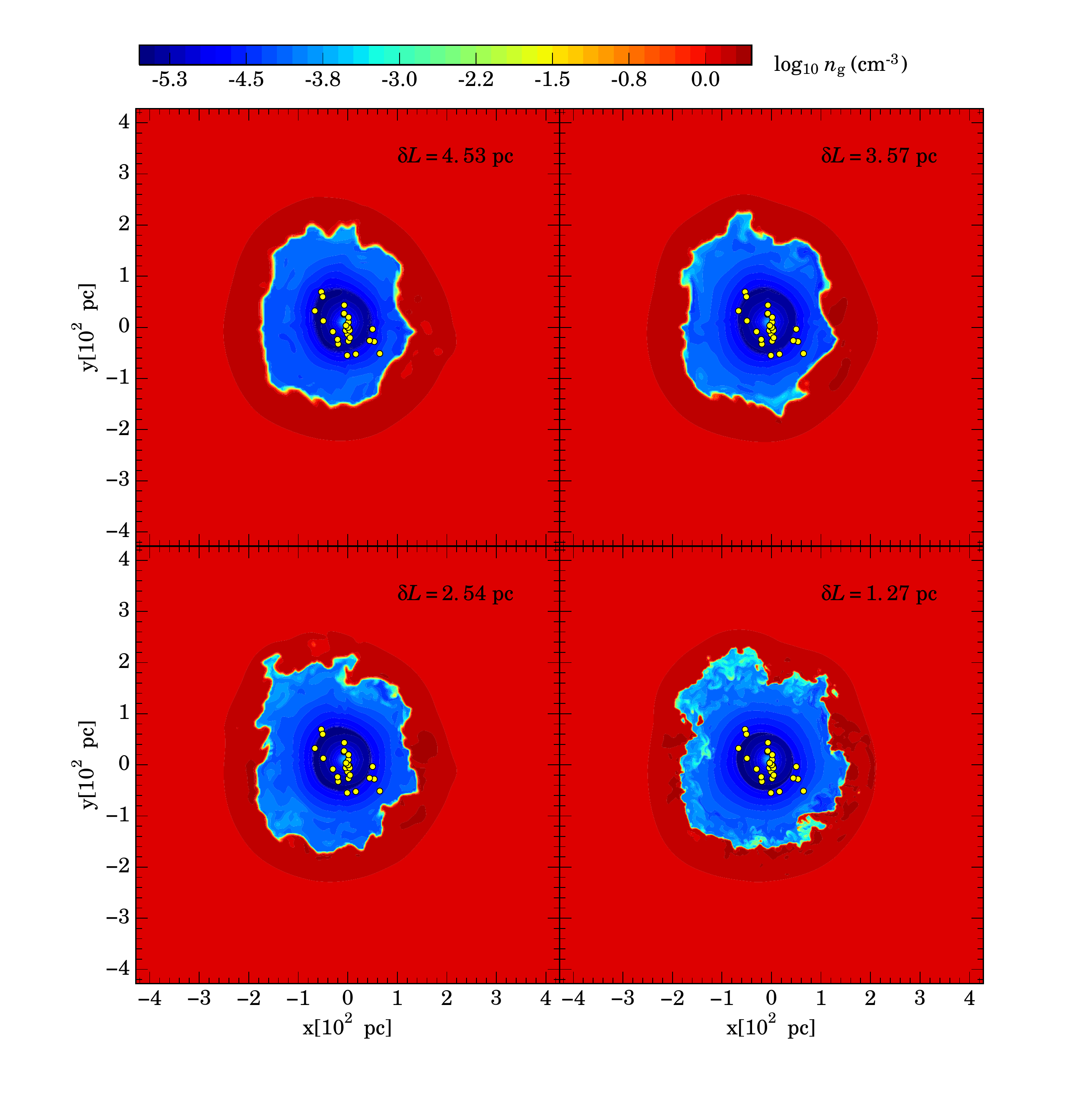}
\caption{Density snapshots in the $x-y$ plane at $9.55\ \rm Myr$ for the fiducial
parameters but with various grid resolutions (see Table~\ref{table:convergence}).
The runs with higher resolution better resolve the features at the bubble-shell
interface. Yellow filled circles indicate the projected locations of SNe
that have gone off.}\label{fig:density-plot-convergence}
\end{figure*}

Fig.~\ref{fig:density-plot-convergence} shows the density snapshots of
four simulations with the grid resolution of $1.27\ \rm pc$, $2.54\ \rm pc$ and
$3.57\ \rm pc$ and $4.53\ \rm pc$ at $9.55\ \rm Myr$. The
simulations with higher resolution better resolve the internal structures
within the bubble. Strict convergence is only expected with explicit viscosity and thermal
conductivity. Since molecular transport is negligible, we do not include these in our simulations.
The run with $\delta L=2.54$ pc looks morphologically very similar to the run with $\delta L=1.27$ pc,
but is $\approx 16$ times faster.
Since simulations of the cluster over its typical
lifetime ($\sim 30 \ \rm Myr$) is computationally expensive, we have chosen a
resolution close to $\delta L \approx 2.54$ (corresponding to run R2.5 in Table~\ref{table:convergence})
for most of our simulations (see Table~\ref{table:simulations}).

\end{document}